\documentclass[onecolumn,trackchanges]{aastex631}

\shorttitle{Understanding the capabilities of TD solar wind simulations}
\shortauthors{Samara et al.}

\usepackage{graphicx}
\usepackage{subfigure}
\usepackage[figuresleft]{rotating}
\usepackage[dvipsnames]{xcolor}
\usepackage[toc,page]{appendix}
\usepackage{academicons}
\definecolor{orcidlogocol}{HTML}{A6CE39}

\hypersetup{colorlinks=true,citecolor=cyan,urlcolor=cyan,linkcolor=blue}
\begin{document}

\title{Why do solar wind models get it wrong: understanding the capabilities of time-dependent solar wind simulations}

\correspondingauthor{Evangelia Samara}
\email{evangelia.samara@nasa.gov, evangelia.sam@gmail.com}

\author[0000-0002-7676-9364]{Evangelia Samara}
\affiliation{NASA Goddard Space Flight Center, Greenbelt, MD, USA}
\affiliation{The Catholic University of America, Washington, D.C., USA}

\author[0000-0001-8875-7478]{Elena Provornikova}
\affiliation{Johns Hopkins University Applied Physics Laboratory, Laurel, MD, USA}

\author[0000-0001-9326-3448]{C. Nick Arge}
\affiliation{NASA Goddard Space Flight Center, Greenbelt, MD, USA}

\author[0000-0002-6222-3627]{Andrew J. McCubbin}
\affiliation{Johns Hopkins University Applied Physics Laboratory, Laurel, MD, USA}

\author[0000-0003-4344-5424]{Viacheslav G. Merkin}
\affiliation{Johns Hopkins University Applied Physics Laboratory, Laurel, MD, USA}

\begin{abstract}

{We explore the capabilities of time-dependent (TD) magnetohydrodynamics (MHD) solar wind simulations with the coupled WSA model of the solar corona and GAMERA model of the inner heliosphere. We compare TD with steady state (SS) simulations and with in situ data from multiple spacecraft (Earth, STEREO-A, PSP). We show that TD predictions, although better than SS predictions, substantially mispredict the solar wind at different heliospheric locations. We identified three reasons for that: (1) the uncalibrated WSA velocity formula used to generate solar wind velocities at the inner boundary of a heliospheric domain, (2) the extraction of the WSA boundary conditions for input into MHD models very high in the corona, and (3) the abrupt and partial emergence of active regions from the solar east limb. Evaluation of one year of TD predictions at the Earth and STEREO-A locations shows that by tuning accordingly the WSA relationship when used with MHD models, and by extracting the WSA boundary conditions lower in the corona (at 5~Rs instead of 21.5~Rs), can lead to improved predictions. However, the abrupt emergence of active regions from the east limb of the Sun which can highly disrupt the magnetic field topology in the corona, is a difficult task to deal with since complete knowledge of the conditions on the solar far side is not currently available. Solar Orbiter Polarimetric and Helioseismic Imager (PHI) data can help mitigate this effect, however, unless we get a 4$\pi$ view of the Sun we will be unable to completely address it.}

\end{abstract}


\section{Introduction} \label{Sec:Introduction}

It has been more than twelve years since the concept of time-dependent (TD) magnetohydrodynamic (MHD) solar wind simulations in the inner heliosphere was introduced to the heliophysics community. The term ``time-dependent" here refers to the TD treatment of the boundary conditions at the inner boundary of heliospheric models (typically at 21.5~Rs) that are necessary to drive a dynamic solar wind. This is achieved by employing a series of global photospheric magnetic field maps that, when inserted in a coronal model, help us approximate the evolution of the solar corona and generate the necessary updated solar wind plasma and magnetic field conditions for the driving of the solar wind in the heliospheric MHD domain.
Such approach is assumed to be more accurate compared to the traditional steady state (SS) approach for which we usually employ a single magnetogram and a single set of boundary conditions produced at a particular moment (i.e., snapshot of the solar corona) to drive heliospheric models for a range of days or weeks \citep[see e.g.,][]{Linker2016, merkin2016time, owens2024}. 

One of the first studies that presented such TD approach was from \cite{hayashi2012}. In his study, they used Wilcox Solar Observatory (WSO) magnetograms \citep[][]{hoeksema1986} and interplanetary scintillation (IPS) observations \citep[][]{tokumaru2010} to specify the heliospheric inner boundary conditions at approximately 50~Rs in order to drive the solar wind until 6~AU. They performed simulations for nine Carrington Rotations (CR; CR numbers 1841 to 1849) ranging from May to December 1991. During this time interval, they updated the boundary conditions once each CR in an effort to determine a more realistic MHD state of the solar wind at distances far away from the Sun. Comparison of their modeling results with data at Earth and Ulysses showed overall good agreement, however, TD predictions were not very different from SS ones most probably due to the very infrequent update of the photospheric maps. They also showed that some magnetic field lines lose connectivity to the inner boundary sphere, mostly due to numerical diffusion issues. 

In 2016 \cite{linker2016empirically} by using full disk magnetograms obtained from the National Solar Observatory Synoptic Optical Long-term Investigations of the Sun (NSO SOLIS) Vector Spectromagnetograph (VSM) combined with the Assimilative Photospheric flux Transport \cite[ADAPT;][]{arge2010air, arge2011, arge2013, henney2012forecasting, henney2015, hickmann2015} model, showed the dynamic evolution of the solar wind in the heliospheric domain of the Magnetohydrodynamic Algorithm outside a Sphere \citep[MAS;][]{Riley2011} MHD model. MAS was driven by TD boundary conditions at 30~Rs based on the combination of the empirical potential field source surface \citep[PFSS;][]{altschuler1969magnetic_npp,schatten69,Wiegelmann2017} model, the potential field current sheet \citep[PFSC;][]{schatten1971current_npp} model and the distance to the coronal hole boundary \citep[DCHB;][]{riley2001} model. The authors updated the magnetograms daily for a period of one year (09/2003 to 09/2004). They observed features that were not found in SS simulations, such as the emergence of evolving fast streams, and they also located magnetic field lines that seemed disconnected from the inner boundary of the heliospheric domain most probably due to changes in the heliospheric current sheet from one magnetogram to another.

The same year, \cite{merkin2016time} also performed TD solar wind simulations with the Lyon-Fedder-Mobarry \citep[LFM;][]{lyon2004} inner heliospheric MHD model, by employing Global
Oscillation Network Group (GONG) ADAPT magnetograms and the Wang-Sheeley-Arge \citep[WSA;][]{ArgePizzo2000, arge03, Arge04} model to generate the necessary TD boundary conditions at 21.5~Rs. They showed that, for a period of two CRs and by updating the photospheric maps daily, TD solar wind simulations reconstruct better the conditions in the inner heliosphere compared to the traditional SS approach. They also thoroughly discussed the deviation of the magnetic field lines from the ideal Parker's spiral shape due to plasma parcels of different speeds propagating on top of the same magnetic field lines as the boundary conditions change. This is a kinematic phenomenon, solely observed in the TD approach because of the frozen-in nature of the magnetic field lines which follow the movement of the plasma.


More recently, \cite{owens2024} used the WSA coronal model with GONG ADAPT magnetograms to drive the TD Heliospheric Upwind eXtrapolation model \citep[HUXt;][]{owens2020computationallyEfficient, barnard2022huxt}. HUXt treats the solar wind as a 1D incompressible hydrodynamic flow and it is used as a surrogate model for 3D MHD processes. The authors compared results of SS and TD approach based on daily updated photospheric maps and they showed that a number of problems occurring in the solar wind forecasting due to the SS approaches are alleviated with the TD method. Such problems have to do with the fact that (a) forecasts based on observations from 3-days previous are more accurate than forecasts based on the most recent observations, (b) forecasts for a given day can jump significantly as new observations become available, changing CME propagation times by up to 17~h, and (c) the fact that the heliospheric magnetic field, which controls the solar energetic particle propagation to Earth, can stay highly invariable, opposite to what we see in observations. Other recent works with TD solar wind MHD simulations are those from \citet[][]{odstrcil2023TD} who showed new features of the WSA-ENLIL-Cone modeling system which can simulate continuously evolving background solar wind and multi-CMEs events observed by various spacecraft as well as \citet[][]{baratashvili2025TD} who upgraded the capabilities of the Icarus model \citep[][]{verbeke2022, baratashvili2022} to dynamic solar wind modeling.

In this work, we present for the first time TD solar wind simulations with the Grid Agnostic MHD for Extended Research Applications \citep[GAMERA;][]{zhang2019GAMERA} 3D MHD model of the inner heliosphere driven by the WSA coronal model. Previous works with GAMERA have focused on SS solar wind simulations by either using different flux transport models \citep[][]{knizhnik2024}, or investigating Kelvin–Helmholtz instabilities by using a highly refined spatial grid \citep[][]{mostafavi2022} or launching the Gibson-Low flux rope CME model in the heliospheric domain of GAMERA \citep[][]{provornikova2024}. 
The goal of this paper is to show the TD capabilities of the coupled \mbox{WSA-GAMERA} pipeline in reconstructing the solar wind signatures at multi-spacecraft locations in the inner heliosphere, but also to indicate the extent to which such simulations are successful compared to observations and SS simulations. We emphasize on the fact that despite achieving more accurate predictions compared to SS runs, TD simulations are far from perfect. We investigate in detail the reasons of discrepancies between TD solar wind predictions and observations which are mostly relevant to the coronal boundary conditions that drive them. 
We quantitatively evaluate our model at different spacecraft locations in the heliosphere for a period of one year and show what kind of adjustments on the coronal part of the pipeline can be made for better model-data agreement in the heliosphere. Our study is not only limited to the specific pipeline of models (\mbox{WSA-GAMERA}) but to all modeling pipelines that are using the same principles (namely, a combination of a semi-empirical coronal model, like WSA, and a 3D MHD heliospheric model, like GAMERA). Therefore, we aim to highlight that the major source of uncertainties in solar wind predictions is not so much the details of the solar wind models themselves, but the coronal boundary conditions they are provided. Our analysis and results can therefore be used as guidelines for other modelers, as well.

In section~\ref{Section:Methods} we present the models which we use to perform the MHD simulations and the approach to drive the TD boundary conditions at 21.5~Rs. In section~\ref{Section:Why do models get it wrong} we analyze three different reasons why TD simulations can give us faulty predictions at various points in the heliosphere by examining observed and modeled time series from one Carrington Rotation. Based on our findings, we optimize a number of parameters in the \mbox{WSA-GAMERA} pipeline, and in section~\ref{Evaluation2021} we assess one year of TD predictions from the default and optimized versions according to traditional error functions and the dynamic time warping technique. Evaluation of one year of SS predictions with the optimized modeling set-up, also takes place. Section~\ref{section:Summary&Conclusions} summarizes our results and presents our main conclusions.

\section{Methods}
\label{Section:Methods}

The GAMERA model of the inner heliosphere \citep[][]{zhang2019GAMERA, mostafavi2021, provornikova2024} is a re-invention of the high-heritage LFM code. It was built from scratch in order to improve the LFM numerics and prepare for next-generation supercomputers. The numerical algorithms underlying GAMERA are characterized by extremely low numerical dissipation, the ability to resolve structure within a very limited number (1-2) of cells, and the intrinsic divergence free $\nabla \cdot B$ = 0 update. This is achieved by the use of high-order spatial reconstruction (typically, 7th or 8th order), an aggressive flux limiter, constrained transport, and arbitrary non-orthogonal grids that can be adapted to the problem at hand. GAMERA brings state-of-the-art innovations (e.g., modern software design methods, multiple layers of heterogeneous parallelism, e.g., MPI, OpenMP and SIMD) to the field of heliospheric simulations. 

GAMERA is driven by the WSA model which is a semi-empirical coronal model that provides radial velocity and magnetic field as a function of latitude and longitude at the inner boundary of GAMERA (21.5~Rs). WSA reconstructs the magnetically dominated solar corona by employing the PFSS and the Schatten Current Sheet \citep[SCS;][]{schatten1971current_npp} models. Once the magnetic field line topology has been determined according to those models, the solar wind velocity is estimated at 21.5~Rs based on a parameterized relationship which takes into account the expansion of the magnetic field lines and the proximity of their footpoint to the nearest coronal hole boundary \citep[][]{arge03, Arge04, owens08, mcgregor11, Samara2021}. The WSA model requires quality estimates of the global solar photospheric magnetic field distribution as input. For that, the ADAPT model is often employed, which utilizes the photospheric flux transport framework and advanced data assimilation techniques, taking into account both model and observational uncertainties. ADAPT evolves the global magnetic flux distribution from GONG magnetograms using transport processes when measurements are not available \citep[][]{wordenharvey2000, schrijverDeRosa2003}. It produces significantly more realistic estimates of the instantaneous global photospheric magnetic field distribution than those provided by traditional photospheric field synoptic maps. To complete the set of boundary conditions for GAMERA, density and temperature are typically obtained using empirical relations or assumptions \citep[][]{shiota2014, riley2001}. More specifically, to determine density, an empirical fit to Helios data is used, while temperature is derived assuming the total pressure balance in the angular direction \citep[][]{merkin2016}.

In addition to the \mbox{WSA-GAMERA} pipeline, we use the full WSA 1D kinematic code (from now on it will be mentioned as ``WSA") based on which the velocities that are generated at the outer boundary of its coronal domain (at 5~Rs, in this case), are propagated outwards according to a 1D ballistic approach \citep[][]{ArgePizzo2000, arge03, Arge04, Owens05}. The velocity of the solar wind is radial and does not change with distance except to conserve mass and mass flux at the stream interaction regions. No acceleration of the solar wind takes place as it propagates outwards, only decelerations at the stream interaction regions. Our results with \mbox{WSA-GAMERA} will be compared to the ones from WSA as the latter model is routinely used for decades within the space weather community and can help us explain discrepancies with the \mbox{WSA-GAMERA} output when the two of them show very different results.

Following the methodology of \citet[][]{merkin2016time}, TD simulations with \mbox{WSA-GAMERA} are performed by daily updating the magnetograms and thus the WSA boundary conditions at 21.5~Rs. The spatial resolution of the grid used is uniform: (256 × 128 × 256) in (r, $\theta$, $\phi$) directions, respectively. Driving the pipeline with a sequence of photospheric maps provides a good approximation of the evolution of the solar corona and, as a result, a good approximation of the evolving solar wind in the inner heliosphere. To evolve the boundary conditions from one timestamp to another, we perform linear interpolation of the plasma and magnetic parameters (radial velocity, density, temperature, radial magnetic field) at 21.5~Rs. Specifically for the magnetic field, GAMERA uses the constrained transport to ensure $\nabla \cdot B$ = 0, therefore, Br is evolved by applying the corresponding calculated tangential electric field components, as described in detail in \citet[][]{merkin2016time}. 

\section{Why do solar wind models get it wrong: exploring factors that contribute to prediction discrepancies} 
\label{Section:Why do models get it wrong}

In the left panel of Figure~\ref{Fig:STA_Sept2021} we present the WSA and \mbox{WSA-GAMERA} model results at STEREO-A during September 2021. From top to bottom, the solar wind bulk speed, proton density, temperature, and radial magnetic field are shown as a function of time. The black color represents the combined merged hourly plasma and magnetic field (COHO1HR) observations obtained from the NASA's Coordinated Data Analysis Web (CDAWeb; \url{https://cdaweb.gsfc.nasa.gov}), and the orange dots the WSA three-day advance predictions\footnote{Three-day advance predictions mean that the solar wind at STEREO-A's location has been predicted based on an input map three-days earlier from the date of the solar wind's arrival at STEREO-A.}. The red color is indicative of the SS \mbox{WSA-GAMERA} output performed with the GONG ADAPT photospheric field map of 2021-09-11. The blue color shows the TD \mbox{WSA-GAMERA} output based on GONG ADAPT photospheric maps which were daily updated. For all runs, the first GONG ADAPT realization was considered. 

The first thing we notice is that \mbox{WSA-GAMERA} predictions (red or blue time series) do not reconstruct the observed speeds optimally as they fail to capture the slow solar wind (see period between August~29 - September~9,~2021), the transitions between the slow and fast solar wind (see period between September~15-21,~2021), or they mistakenly predict fast flows (see period between September~21-30,~2021 for red time series and September~27-October~2,~2021 for blue time series). On the other hand, WSA (orange dots) succeeds in capturing most of these large scale characteristics. The question is, what are the factors that contribute to the discrepancies of the \mbox{WSA-GAMERA} output which relies on a more physics based (MHD) approach, compared to the more simplistic approach of the WSA model?  

\subsection{Factor 1: The calibration of the WSA velocity formula at the inner boundary of an MHD heliospheric domain} 

To answer the aforementioned question, we need to remind ourselves that GAMERA (and other inner heliospheric models such as Enlil \cite[][]{Odstrcil04}, EUHFORIA \cite[][]{pomoell18}, etc) are usually driven by the coronal counterpart of the WSA model which relies on the PFSS and SCS models to reconstruct the magnetically dominated corona and produce the solar wind boundary conditions at 21.5~Rs. In our case, the WSA coronal model provides the radial magnetic field, which is calculated at 21.5~Rs according to the PFSS and SCS models, and the solar wind radial velocities at the same heliocentric distance, which are produced based on the WSA semi-empirical equation of the following form: 

\begin{equation}
 \text{v}_{r}(f,d)= V_{0} +\frac{V_{1}}{(1+f)^{\alpha}}\left[1-0.8\exp\left(-\left(\frac{d}{w}\right)^{\beta}\right)\right]^3.
\label{WSA_vr}
\end{equation}

\noindent where $V_{0} =\;$285 km/s, $V_{1} = 625\;$km/s, $\alpha$ = 0.222, $\beta$ = 2 and $w = 2\;$rad. However, we should not forget that eq.~\ref{WSA_vr} has been calibrated in the above-mentioned form in order to give well-modeled speeds at Earth according to the 1D kinematic (ballistic) approach and a specific type of global photospheric magnetic field maps (originally Mount Wilson Solar observatory maps and more recently, GONG and VSM maps). \cite{mcgregor11, Samara2024} have tried to calibrate this relationship according to the Helios \citep[][]{helios_mission} and Parker Solar Probe \citep[PSP;][]{fox2016, PSPmission2, PSPmission1, raouafi2023parker} data, respectively, for the WSA-Enlil and WSA-EUHFORIA models. Their calibrations were relevant to the fine-tuning of some of the constant parameters of eq.~\ref{WSA_vr}, which could be improved according to observational data and past studies. In the case of the \mbox{WSA-GAMERA}, eq.~\ref{WSA_vr} has not been calibrated, so we should not necessarily expect the predictions at Earth (or at any other point in the inner heliosphere) to be optimal. 

The main characteristic of the \mbox{WSA-GAMERA} time-series from both the SS and TD approach is the lack of the solar wind bimodality. Not only \mbox{WSA-GAMERA} simulations overestimate the slow solar wind, but they also miss the proper reconstruction of the stream interaction regions (SIRs; where the fast solar wind overtakes the preceding slow solar wind). To account for these discrepancies, we follow the example of \cite{Samara2024}. Namely, we first decrease the value of $V_{0}$ from 265~km/s to 207~km/s in order to capture the slow solar wind better. The parameter $V_{0}$ should always reflect the minimum velocity of the solar wind at the heliocentric distance of interest (in our case, at the boundary of 21.5~Rs). As recorded by PSP during its first eight encounters, this is 207~km/s which is 58~km/s lower than the initial value. Also, in order to improve the dynamics and the transitions between the slow and fast solar wind, we increase the value of $\beta$ from 2 to 3. The parameter $\beta$ defines how abrupt the transition from slow to fast solar wind is at the coronal hole boundary on the photosphere, therefore, a higher value of it reflects a sharper transition \citep[see][for more details]{Samara2024}. Based on the aforementioned modifications, the optimized WSA velocity relationship takes the following form: 

\begin{equation}
 \text{v}_{r}(f,d)= 207 +\frac{700}{(1+f)^{\frac{2}{9}}}\left[1-0.8\exp\left(-\left(\frac{d}{2}\right)^{3}\right)\right]^3.
\label{WSA_vr_optimized}
\end{equation}

\begin{figure}[h!]
\centering
\gridline{\fig{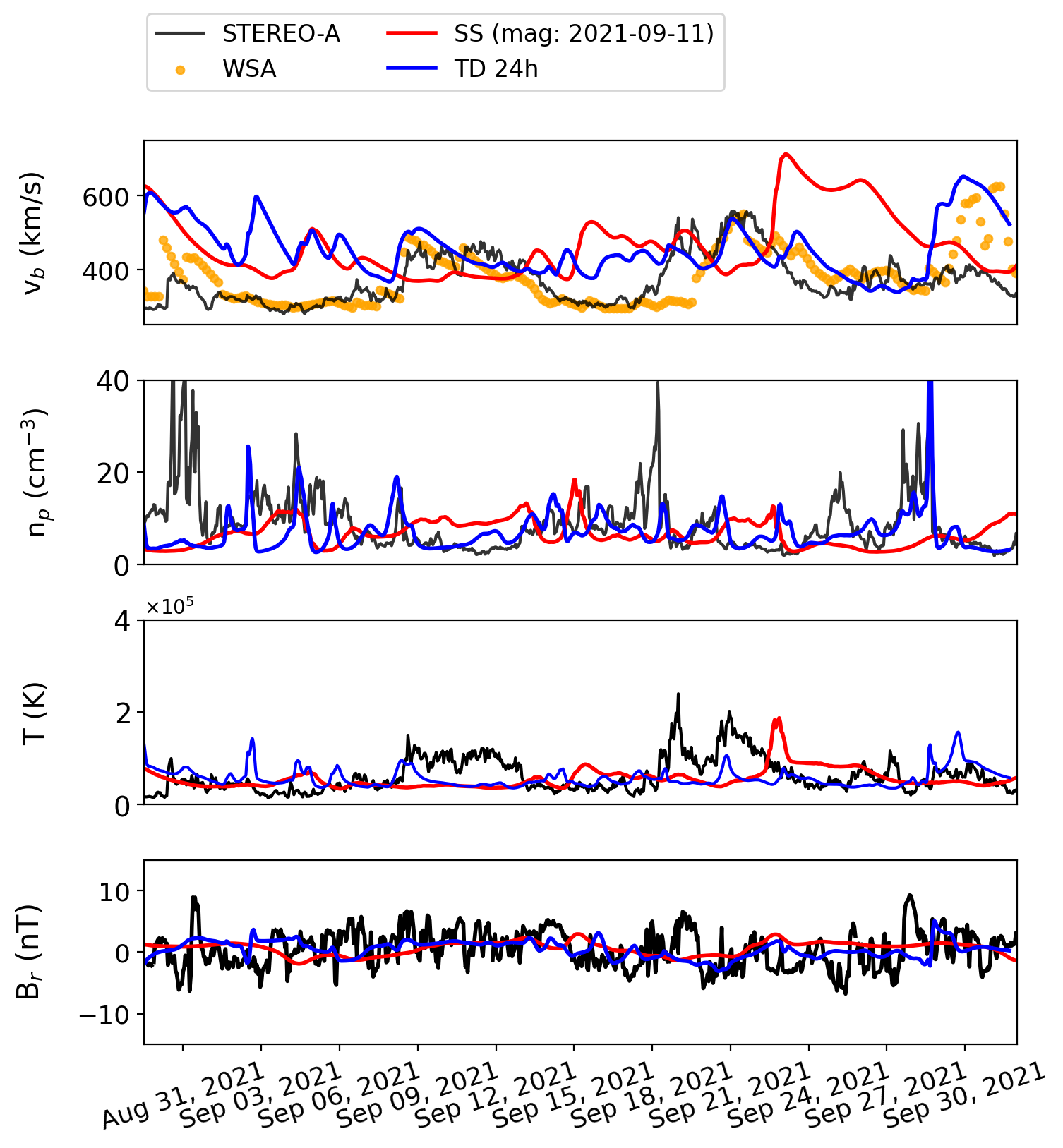}{0.5\textwidth}{}
            \fig{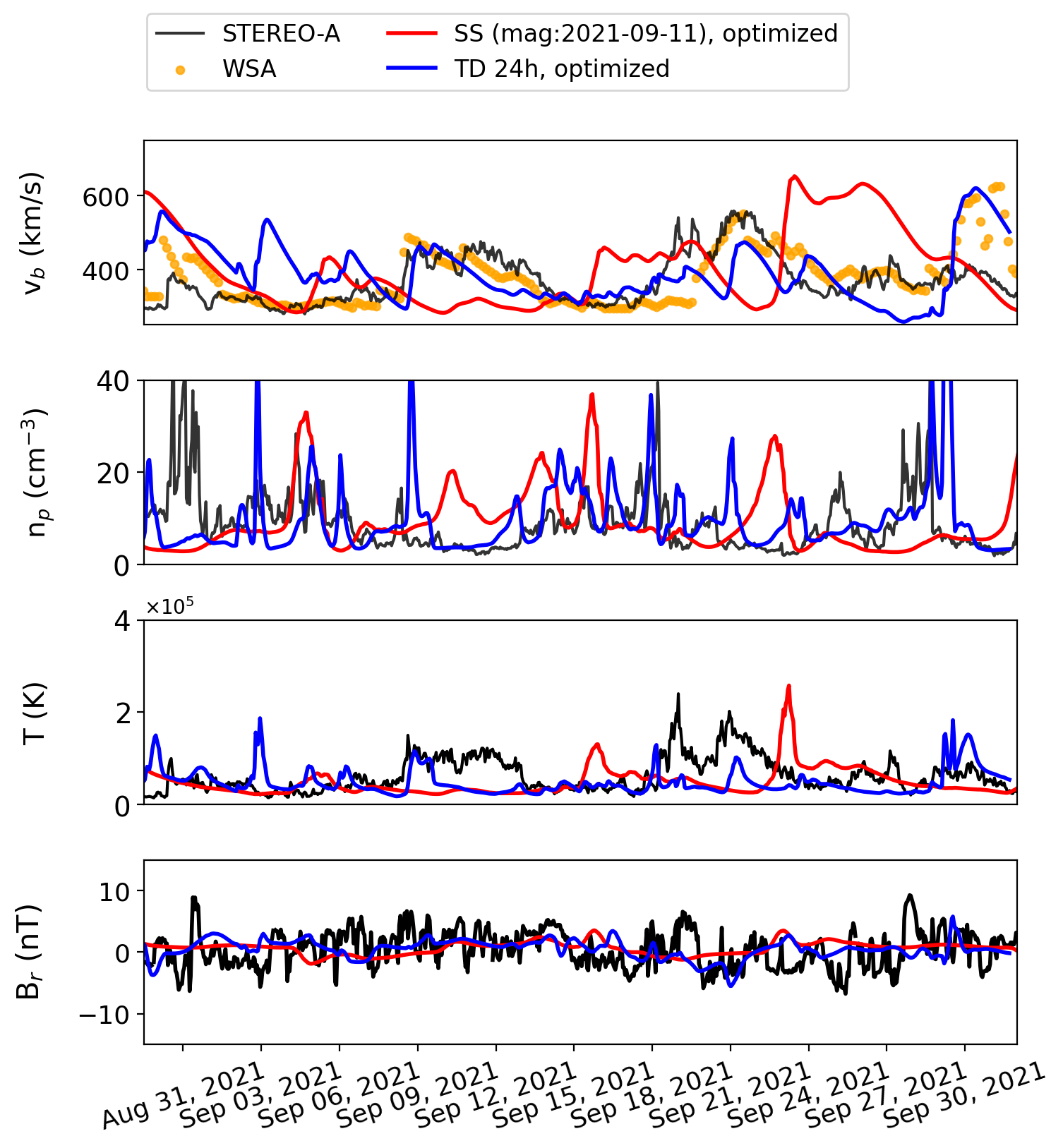}{0.5\textwidth}{}
          }
           
\caption{Comparison of in situ solar wind observations (black) with predictions from WSA (orange dots) and \mbox{WSA-GAMERA}, at STEREO-A. The red color is indicative of the SS \mbox{WSA-GAMERA} simulations performed with the first realization of the GONG ADAPT magnetogram on 2021-09-11. The blue color shows the TD \mbox{WSA-GAMERA} simulations as we update the input GONG ADAPT magnetograms daily. Left panel: SS and TD predictions with \mbox{WSA-GAMERA} based on the default WSA velocity formula (eq.~\ref{WSA_vr}). Right panel: SS and TD predictions with \mbox{WSA-GAMERA} based on the optimized WSA velocity formula (eq.~\ref{WSA_vr_optimized}).}
\label{Fig:STA_Sept2021}
\end{figure}

The updated, optimized time series for the SS and TD approach are seen in the right panel of Figure~\ref{Fig:STA_Sept2021}. We notice that solar wind speeds are not as overestimated as they are in the left panel. According to the TD solar wind simulations in particular, we capture better the slow solar wind during the period September~14-19,~2021, as well as the transitions from slow to fast wind around September~9,~2021 and September~18-24,~2021. On the other hand, the quality of the SS predictions is far from optimal since we used a single input map (on 2021-09-11) to simulate the whole month of September~2021. The SS approach falsely predicts high speed streams (HSSs) where there are none (e.g., September~23-30,~2021), or entirely misses the arrival of fast flows (e.g., September~9-15,~2021). This is because one map cannot capture the continuously evolving conditions in the solar photosphere and corona, and thus it is not enough to properly reconstruct the correct heliospheric conditions farther out. Therefore, it is evident that a combination of TD approach and proper calibration of the WSA velocity formula is crucial for improving several aspects in our solar wind MHD predictions with \mbox{WSA-GAMERA}. It is also important to note that the WSA predictions (orange dots) remain unchanged between the left and right panel of Figure~\ref{Fig:STA_Sept2021} and are based on the default WSA velocity formula (eq.~\ref{WSA_vr}). This is because eq.~\ref{WSA_vr} is already well calibrated for the ballistic 1D propagation of the solar wind from the Sun outwards.

For the rest of the solar wind plasma and magnetic signatures (i.e., density, temperature, radial magnetic field) we notice that the optimized TD simulations (blue time series in the right panel of Figure~\ref{Fig:STA_Sept2021}) usually overestimate the density peaks recorded before the arrival of the HSS, indicative of the stream interaction regions. This happens because the velocity transitions from slow to fast solar wind have become more abrupt in the optimized \mbox{WSA-GAMERA} output compared to the default one (blue time series in the left panel of Figure~\ref{Fig:STA_Sept2021}) with the density depending on the velocity as $\approx$ $\frac{1}{\text{v}_{r}^{2}}$. On the other hand, the temperature signatures do not change much between the TD default and optimized solutions. The radial magnetic field is weak in all cases (no matter if it is coming from SS or TD default/optimized simulations) mostly due to the well-known problem of the open missing flux, namely, the fact that the magnetic field strength in interplanetary space is often underestimated when inferred from coronal models \citep[see][and references therein]{linker2017open, wang2022magnetograph, arge2024}.

\subsection{Factor 2: the influence of the height at which the WSA boundary conditions are generated} 

Fine-tuning eq.~\ref{WSA_vr} does not seem to be enough for providing the best solar wind predictions for the studied period as there are still time intervals during which the modeled TD speeds strongly deviate from observations (see right panel of Figure~\ref{Fig:STA_Sept2021}). For example, during the period August~30~-~September~9,~2021, STEREO-A observes slow solar wind (less than 400~km/s) which is captured well by WSA. However, not only does the \mbox{WSA-GAMERA} output overestimate the speed of the recorded solar wind, but it seems to falsely predict a number of narrow HSSs. Why is this the case, and why is the output of the two models so different? 

\begin{figure}[h!]
\centering
\gridline{\fig{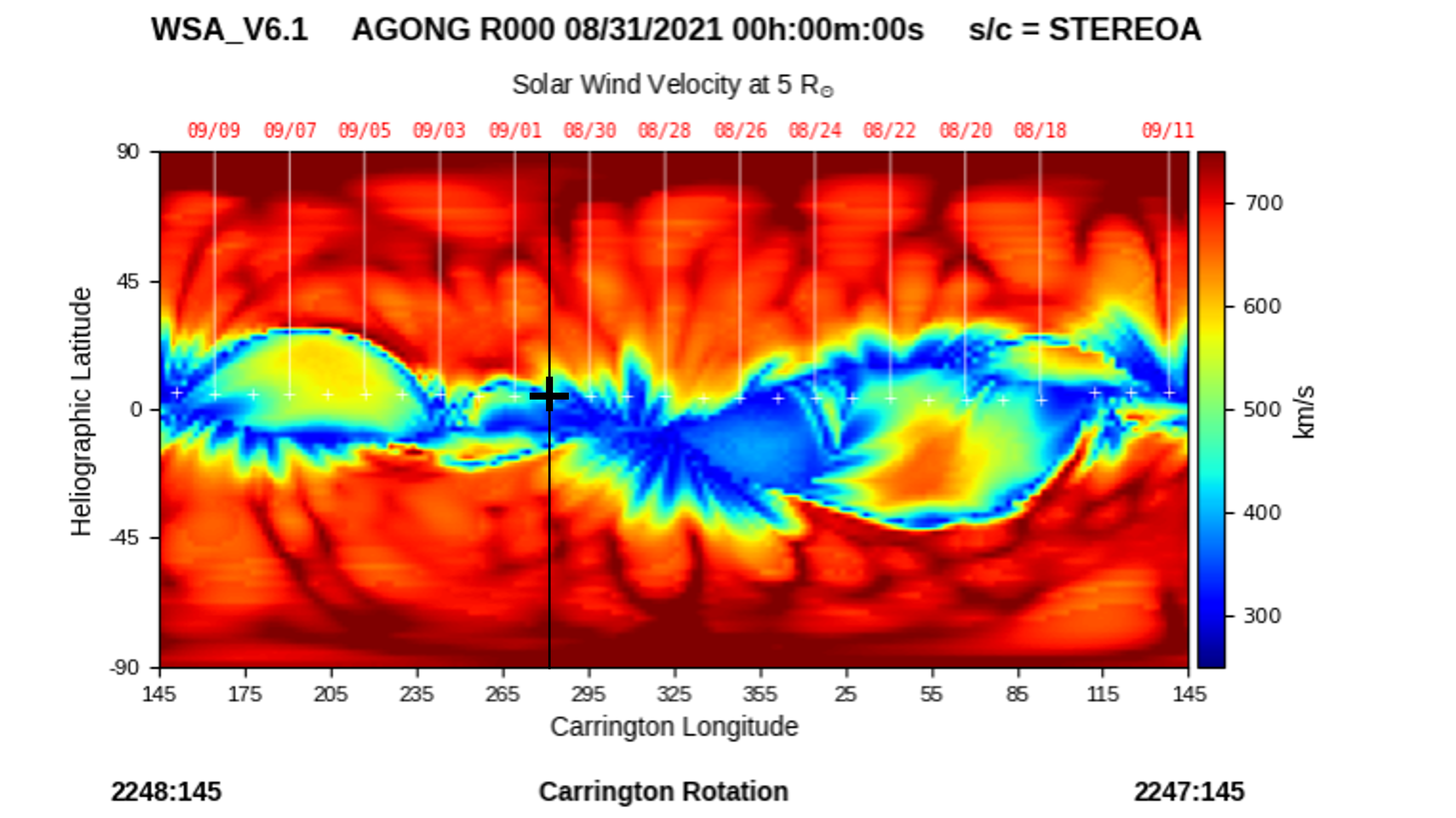}{0.99\textwidth}{}
          }
\gridline{\fig{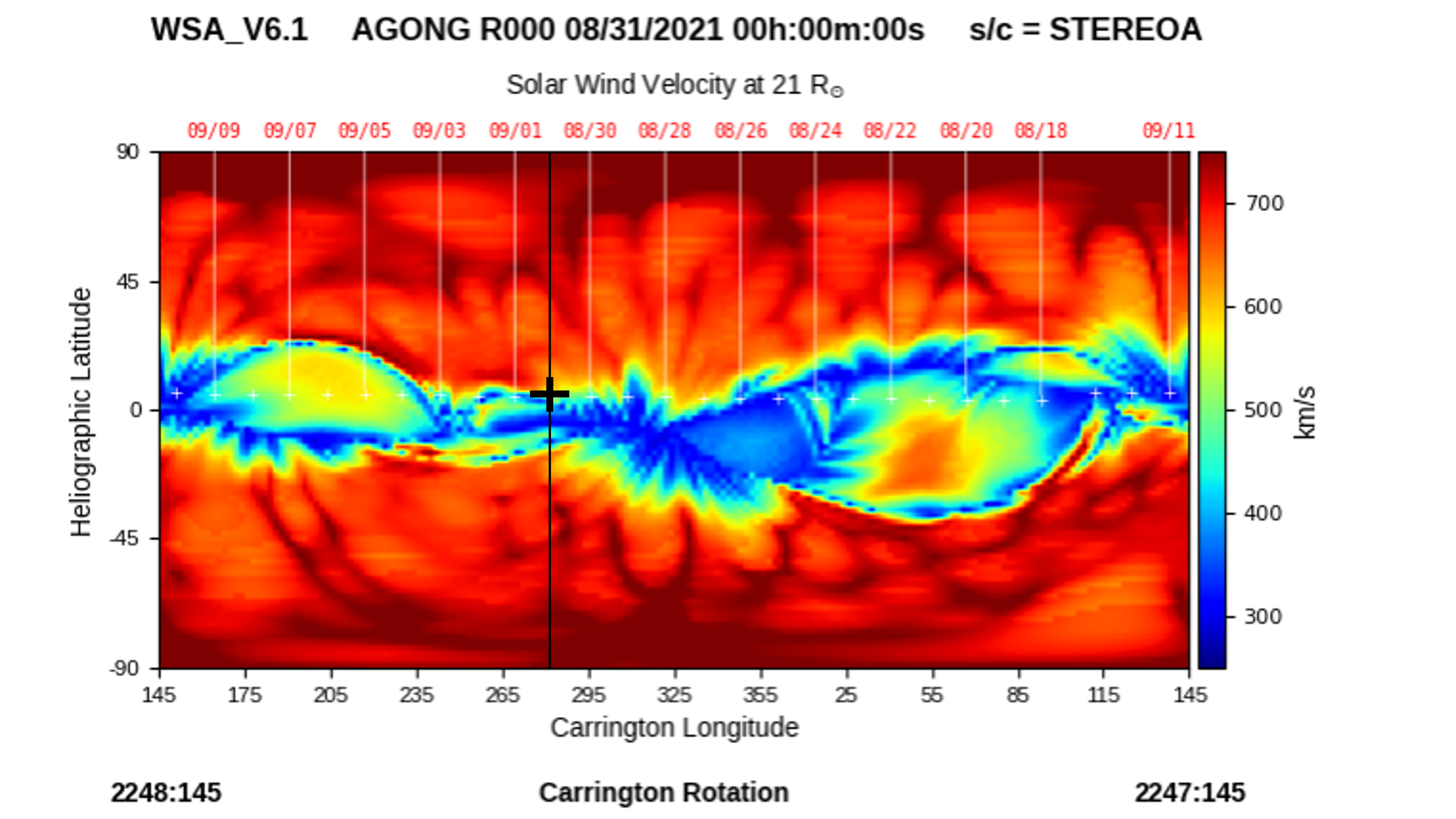}{0.99\textwidth}{}
          }
\caption{WSA solar wind velocity boundary maps at 5~Rs (upper panel) and 21.5~Rs (lower panel). The slowest velocities are shown in dark blue color while the fastest velocities in dark red. The black cross indicates STEREO-A's projection at the outer coronal boundary of 5~Rs (i.e., sub-satellite point) the date that the solar wind left that region (August~31,~2021). Both the black cross and the straight black line have been drawn to facilitate the reader perceiving the HCS flattening from 5~Rs to 21.5~Rs.}
\label{Fig:WSA_boundary_height}
\end{figure}

\begin{figure}
\centering
\gridline{\fig{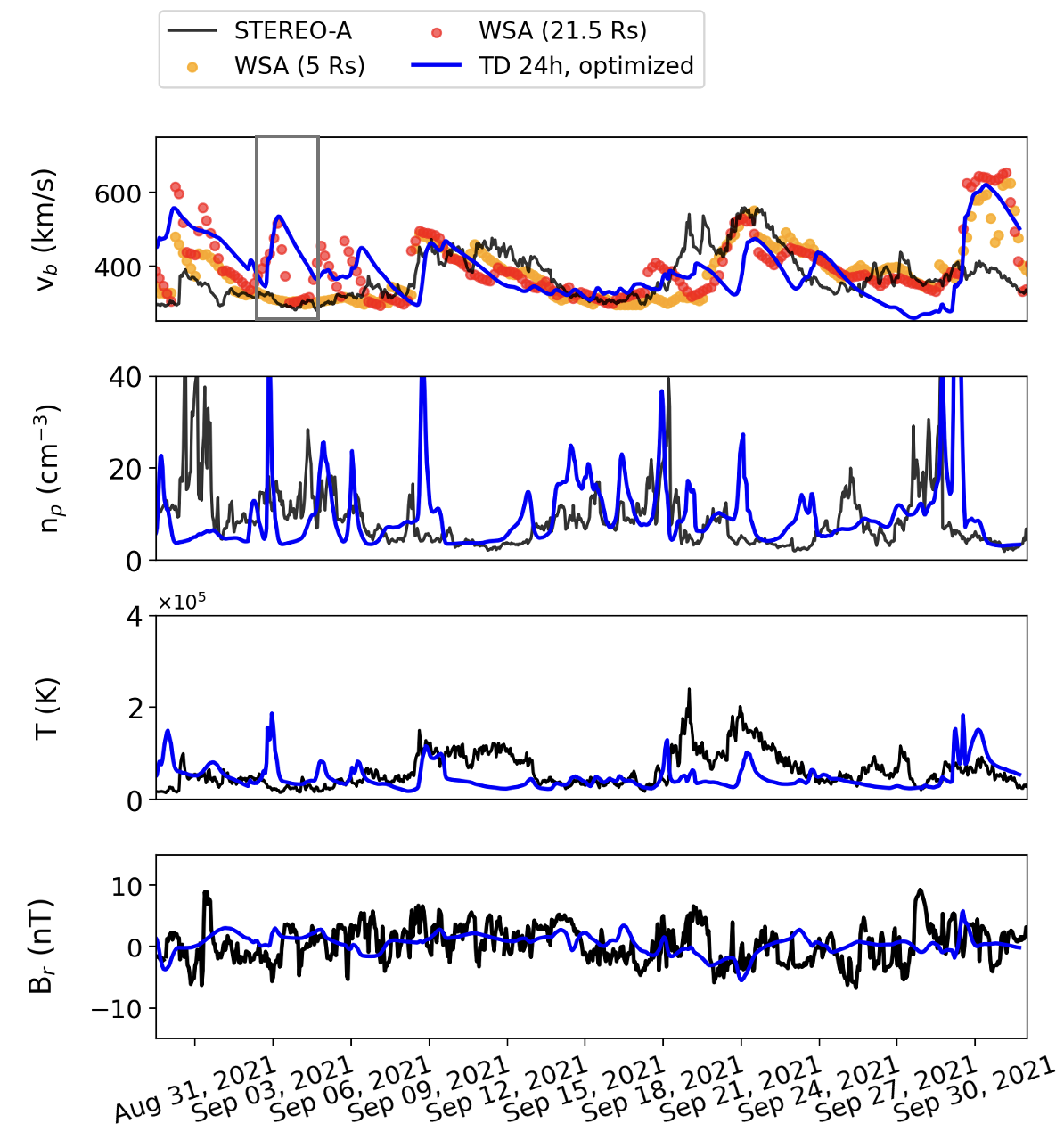}{0.5\textwidth}{}
            \fig{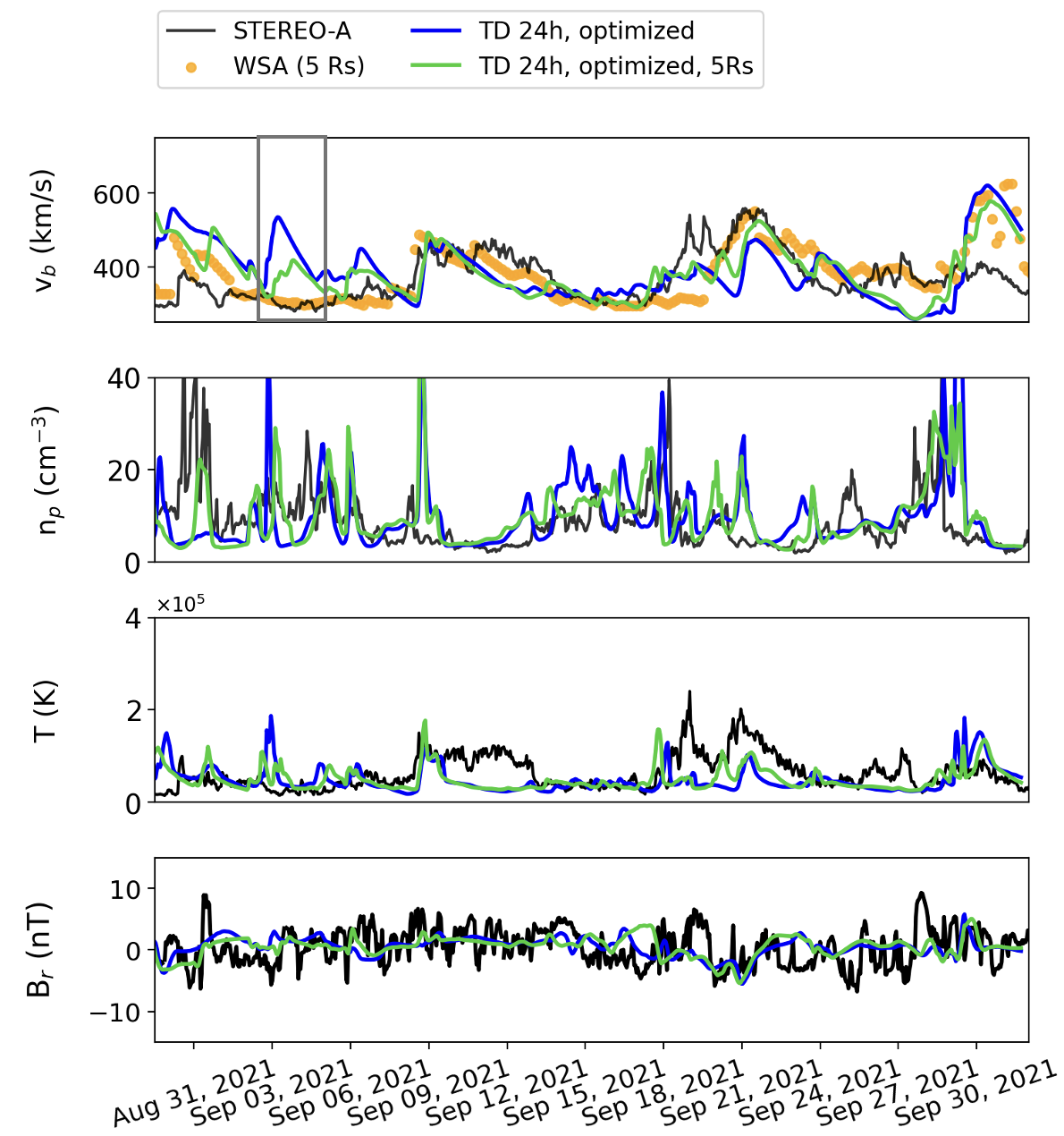}{0.5\textwidth}{}
          }        
\caption{Same as Figure~\ref{Fig:STA_Sept2021} having removed the \mbox{WSA-GAMERA} SS simulations, and with the addition of (a) WSA predictions with the outer coronal boundary at 21.5~Rs (red dots, left panel) and (b) \mbox{WSA-GAMERA} TD solar wind predictions with boundary velocities extracted based on the optimized WSA velocity formula at 5~Rs (green time series, right panel). The gray boxes indicate the arrival of the plasma parcels from the coronal region noted by the black cross in Figure~\ref{Fig:WSA_boundary_height}. The solar wind structure within this box is a good example of how the height at which we extract the WSA boundary conditions can affect the predictions in the inner heliosphere for both the WSA and \mbox{WSA-GAMERA} models.}
\label{Fig:STA_Sept2021_5Rs}
\end{figure}

The answer is given again from the boundary conditions. A major difference between WSA and WSA-GAMERA is where the outer coronal boundary stops. More specifically, the coronal counterpart of the WSA model extends from the solar photosphere until 5~Rs. This is the radius at which the solar wind boundary conditions are produced before they ballistically propagate outward. However, in the \mbox{WSA-GAMERA} model, the outer boundary of the coronal domain extends up to 21.5~Rs. To understand how this difference affects the predictions, we examine Figure~\ref{Fig:WSA_boundary_height}. There, we show the WSA boundary radial velocity map at 5~Rs (upper panel) and 21.5~Rs (lower panel) for the same exact dates and input photospheric map. Dark blue color indicates the slowest velocities while the dark red color shows the fastest velocities. Although the whole coronal structure is qualitatively the same between the two maps, we notice that the region around the heliospheric current sheet (HCS) is flatter when the boundary conditions have been produced at 21.5~Rs, compared to the ones that have been produced at 5~Rs. This can be seen better by observing the straight black line and black cross we have drawn for visual facilitation between the two maps on August~31,~2021. The black cross reveals STEREO-A's projection at the outer coronal boundary of 5~Rs (i.e., sub-satellite point) the date that the solar wind left that region (August 31,~2021). This means that the plasma parcels will reach STEREO-A approximately 2-5 days later (depending on their velocity). Since their velocity here is low ($\approx$~300-400~km/s) the plasma parcels are expected to reach the spacecraft position around September~4-5,~2021. On the other hand, in the lower map of Figure~\ref{Fig:WSA_boundary_height} the sub-satellite point for the same date (black cross) is projected to a region of higher speeds (orange/red color), therefore we should expect speeds of around 500-600~km/s reaching the STEREO-A's location around September~3,~2021. This is confirmed in Figure~\ref{Fig:STA_Sept2021_5Rs}. The left panel of Figure~\ref{Fig:STA_Sept2021_5Rs} is the same as the right panel of Figure~\ref{Fig:STA_Sept2021} having removed the SS simulation and with the addition of the WSA predictions whose boundary conditions were now produced at 21.5~Rs (red dots). We notice that these predictions are indeed much different at the beginning of the studied period compared to the ones that have been created with the outer boundary at 5~Rs (orange dots). They also resemble the \mbox{WSA-GAMERA} predictions, something that confirms the fact that the height of the WSA coronal outer boundary does affect the heliospheric simulations. The time-interval surrounded by the gray box indicates the arrival of the plasma parcels that originated from the region of the black cross in Figure~\ref{Fig:WSA_boundary_height}. Indeed, when the boundary of 21.5~Rs is used by the WSA model, speeds of around 500~km/s arrive at the STEREO-A position on September 3, 2021. When the default 5~Rs WSA boundary is used, speeds of around 300~km/s reach STEREO-A between September 4-5, 2021. 

The height at which the WSA boundary conditions are generated also affects the TD \mbox{WSA-GAMERA} predictions as seen in the right panel of Figure~\ref{Fig:STA_Sept2021_5Rs}. This is the same as the right panel of Figure~\ref{Fig:STA_Sept2021} having removed the SS simulation, and with the addition of the TD \mbox{WSA-GAMERA} predictions (green time series) produced based on WSA radial velocity maps created at 5~Rs instead the usual 21.5~Rs. In doing this, it is important to note two things: (a) we did not change the inner boundary of the GAMERA heliospheric domain from 21.5~Rs to 5~Rs, we just provided WSA velocity maps extracted at 5~Rs instead of 21.5~Rs assuming no acceleration of the solar wind between these two radii, and (b) the radial magnetic field at 5~Rs was scaled accordingly to 21.5~Rs before being fed into GAMERA so that we avoid any sub-sonic and sub-alfvenic velocities entering its heliospheric domain. These simulations (green time series in the right panel of Figure~\ref{Fig:STA_Sept2021_5Rs}) show that the narrow, falsely predicted HSSs between August~31-September~9,~2021 are greatly diminished compared to the simulations created from the WSA boundary conditions based on the optimized velocity formula of eq.~\ref{WSA_vr_optimized} at 21.5~Rs (blue time series in the left and right panel of Figure~\ref{Fig:STA_Sept2021_5Rs}).


The flattening of the HCS detected in the lower map of Figure~\ref{Fig:WSA_boundary_height} as we increase the outer coronal boundary height from 5~Rs to 21.5~Rs is due to the SCS model. More specifically, the purpose of this model is to create a thin structure for the HCS and an approximately uniform coronal magnetic field away from the Sun, consistent with the Ulysses observations. In the WSA model, the SCS counterpart extends from 2.4~Rs to 5~Rs. In the \mbox{WSA-GAMERA} pipeline, it extends from 2.4~Rs to 21.5~Rs so it has more space to act and to create an overly flat structure for the HCS that can obviously affect the heliospheric predictions. This overly flat structure is mostly due to the fact that the gas pressure is ignored in the SCS model while the magnetic pressure dominates, something that does not reflect reality away from the Sun.


\subsection{Factor 3: The abrupt and partial emergence of active regions} 

\begin{figure}
\centering
\gridline{\fig{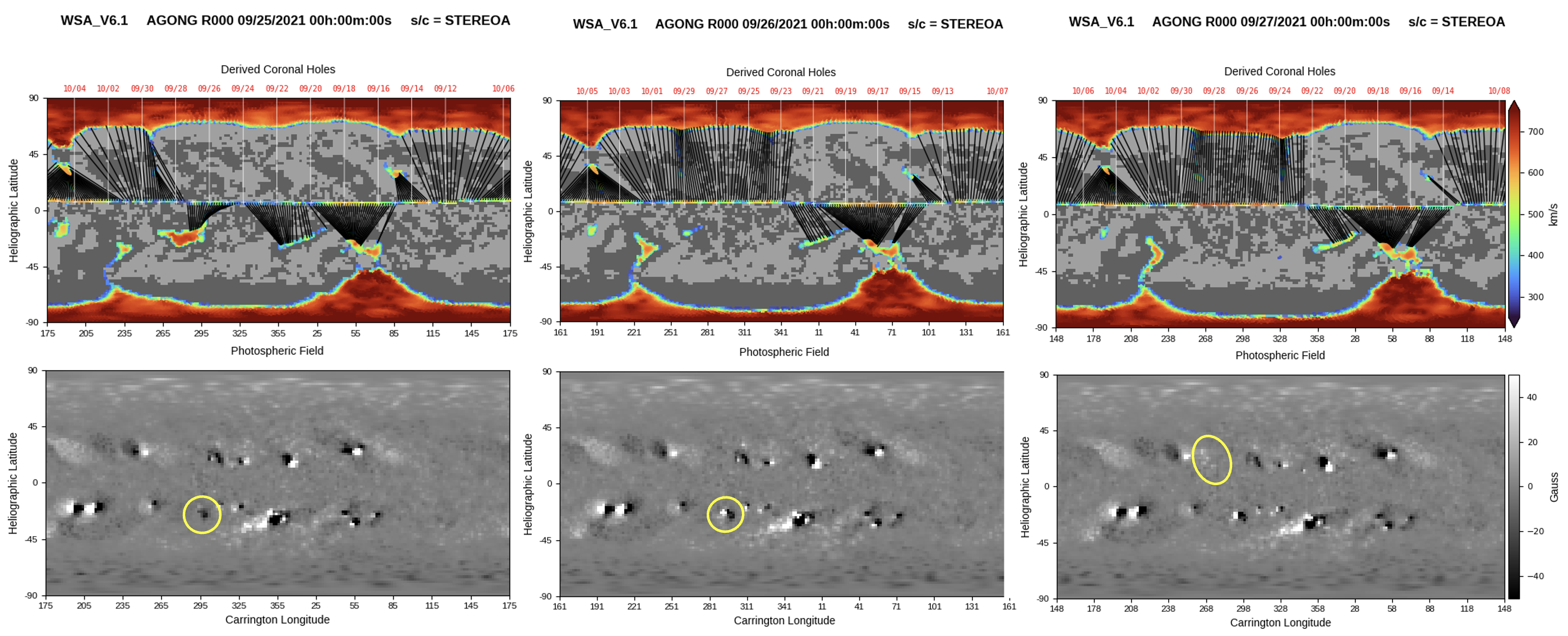}{0.99\textwidth}{}
          }     

\gridline{\fig{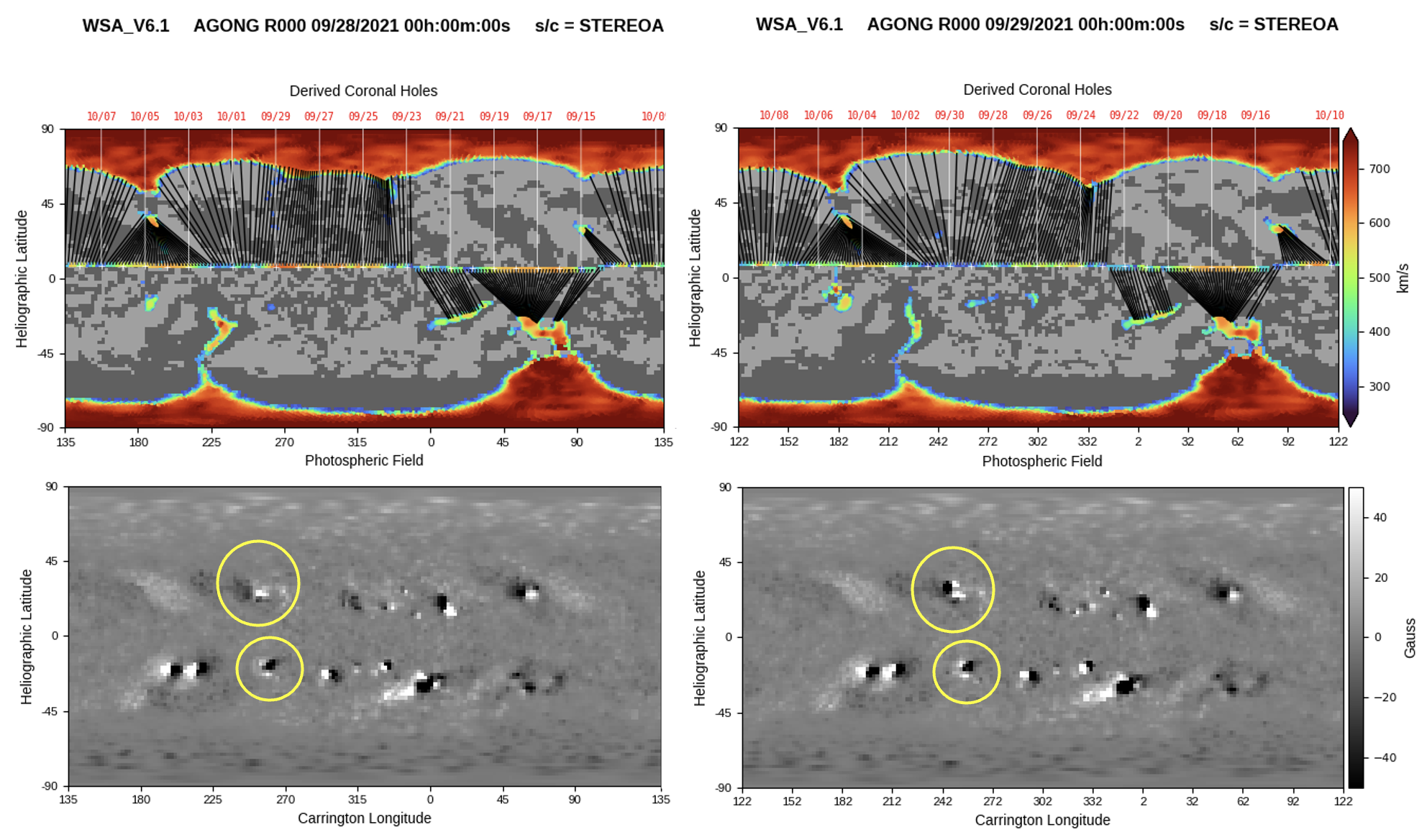}{0.7\textwidth}{}
          }       
\caption{Upper panels: magnetic connectivity between the projection of STEREO-A's location at the outer coronal boundary of 21.5~Rs (i.e., sub-satellite points) and the solar wind source region at 1~Rs, for five consecutive days (September 25 - 29,~2025). Lower panels: same as upper panels but showing the input photospheric magnetic field based on which the connectivity maps were created. Emergence of active regions from the east limb of the Sun from one day to another is circled in yellow.}
\label{Fig:STA_Connectivity}
\end{figure}

As seen in Figure~\ref{Fig:STA_Sept2021_5Rs}, considering the optimized WSA velocity formula and extracting the WSA boundary conditions at a lower heliocentric distance is still not enough to provide ideal predictions from \mbox{WSA-GAMERA}. More specifically, during the time period September~29~-~October~2,~2021, the \mbox{WSA-GAMERA} model predicts an overly fast HSS at the STEREO-A location, with velocities exceeding 600~km/s. The observed speeds during the same period were at least 200~km/s lower. The WSA model also predicts the same HSS no matter if the outer coronal boundary is set to 5~Rs or 21.5~Rs. To understand the reason of the models' discrepancies compared to observations, we study the maps of Figure~\ref{Fig:STA_Connectivity}. The upper panels show with light/dark gray color the positive/negative magnetic field polarity at the photosphere, respectively. The colorful regions indicate the solar wind source regions (coronal holes) at 1~Rs. The color in them (from dark blue to dark red) reflects the slowest and fastest solar wind velocities, respectively. The black lines show the magnetic connectivity between the projection of the spacecrafts's location at 21.5~Rs (i.e., sub-satellite points) and the solar wind source regions at 1~Rs, for five consecutive days (September~25-29,~2025). It is important to note that the spacecraft's connectivity to the sub-satellite points reveals when the solar wind left the Sun as opposed to when it arrived at the spacecraft (e.g., 3-5 days later for 1~AU spacecraft). These maps correspond to the red-dotted WSA predictions of Figure~\ref{Fig:STA_Sept2021_5Rs}, left panel.
The lower panels of Figure~\ref{Fig:STA_Connectivity} show the global input photospheric magnetic field maps used to drive WSA for the same dates, based on which the connectivity maps were created. 

Since both WSA predictions shown in the left panel of Figure~\ref{Fig:STA_Sept2021_5Rs} (orange and red dots) are three-day advance predictions, it means that each plasma parcel at the STEREO-A location has been modeled according to an input map approximately three-days earlier from the date of its arrival. Therefore, the plasma parcels at STEREO-A on September~28,~2021, (when the predicted velocities are still in agreement with observations) are coming from the input map of September 25,~2021, but from where exactly on that map? The answer is given if we consider that the spacecraft's connectivity to the sub-satellite points reveal when the solar wind \textit{left the Sun} as opposed to when it arrived at the spacecraft (e.g., 3-6 days later for 1~au spacecraft). Therefore, for three-day advance forecasts and for plasma parcels of $\approx$ 350~km/s recorded at STEREO-A on September 28,~2021, the solar wind originated from around the sub-satellite point connected to the September 23, 2021 line on the September 25,~2021 map. Similarly, we can identify from which point at the Sun any other plasma parcel of interest originated.

To understand why both WSA and \mbox{WSA-GAMERA} overpredict the solar wind at the end of the studied period, we need to carefully examine the maps of Figure~\ref{Fig:STA_Connectivity}. More specifically, the magnetic field during September 25,~2021 at~00:00UT is characterized by many active regions, one of which seems to be only partially emerged from the east limb as we see its monopolar part (area circled in yellow). The next day (September 26,~2021; see upper middle panels of Figure~\ref{Fig:STA_Connectivity}) we observe the appearance of the rest of the active region (white part appears within the circled yellow area) and the connectivity with the STEREO-A sub-satellite points abruptly changes from the southern, negative polarity coronal hole to the northern polar coronal hole of positive polarity. On September~27,~2021 (see upper right panels of Figure~\ref{Fig:STA_Connectivity}) we observe the appearance of small monopolar regions emerging (also circled in yellow) and the STEREO-A sub-satellite points are connected deeper into the northern polar coronal hole (higher speeds are indicated with orange/red colors around longitude 285$^{o}$). 
By the beginning of September 28,~2021, partial appearance of strong active regions is taking place (see lower left panels of Figure~\ref{Fig:STA_Connectivity}) which is completed by September 29,~2021 (lower right panels of Figure~\ref{Fig:STA_Connectivity}). Therefore, it is clear that the abrupt and partial appearance of active regions between September~26-29,~2021, disrupted the global magnetic field topology, causing faulty connectivities with the STEREO-A sub-satellite points and leading to high solar wind velocities between September~29~-~October~1,~2021. Velocities come back to quiet levels at the end of October~1,~2021, which is in accordance with the full emergence of the active regions and the reinstatement of a more realistic magnetic field topology in the corona on September~29,~2021. This example strongly indicates that unless we get a 4$\pi$ view of the solar disk \citep[see, e.g.,][]{firefly_mission}, we will be unable to improve discrepancies on our predictions that are due to phenomena that are taking place on the far side of the Sun.


\begin{figure}
\centering
\gridline{\fig{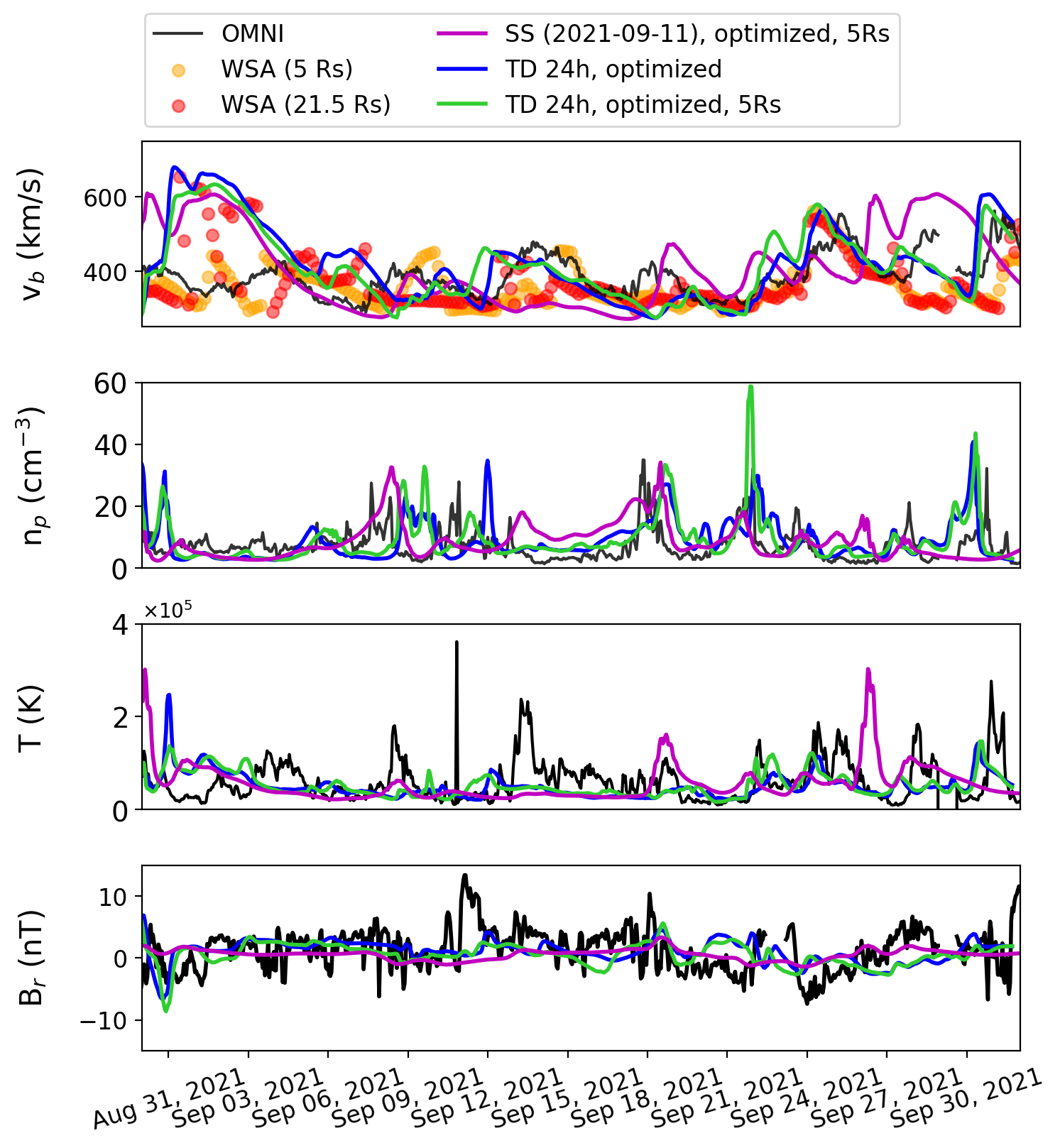}{0.7\textwidth}{}
          }        
\caption{Comparison of in situ solar wind observations (black) with WSA and \mbox{WSA-GAMERA} predictions at Earth. Predictions from WSA having employed the outer coronal boundary at 5~Rs and 21.5~Rs are shown in orange and red dots, respectively. SS simulations from \mbox{WSA-GAMERA} according to the optimized WSA velocity formula of eq.~\ref{WSA_vr_optimized} and WSA boundary maps created at 5~Rs, are shown in magenta. Blue and light green time series are the same as in the right panel of Figure~\ref{Fig:STA_Sept2021_5Rs} but for Earth's location.}
\label{Fig:ACE_Sept2021}
\end{figure}

\begin{figure}
\centering
\gridline{
            \fig{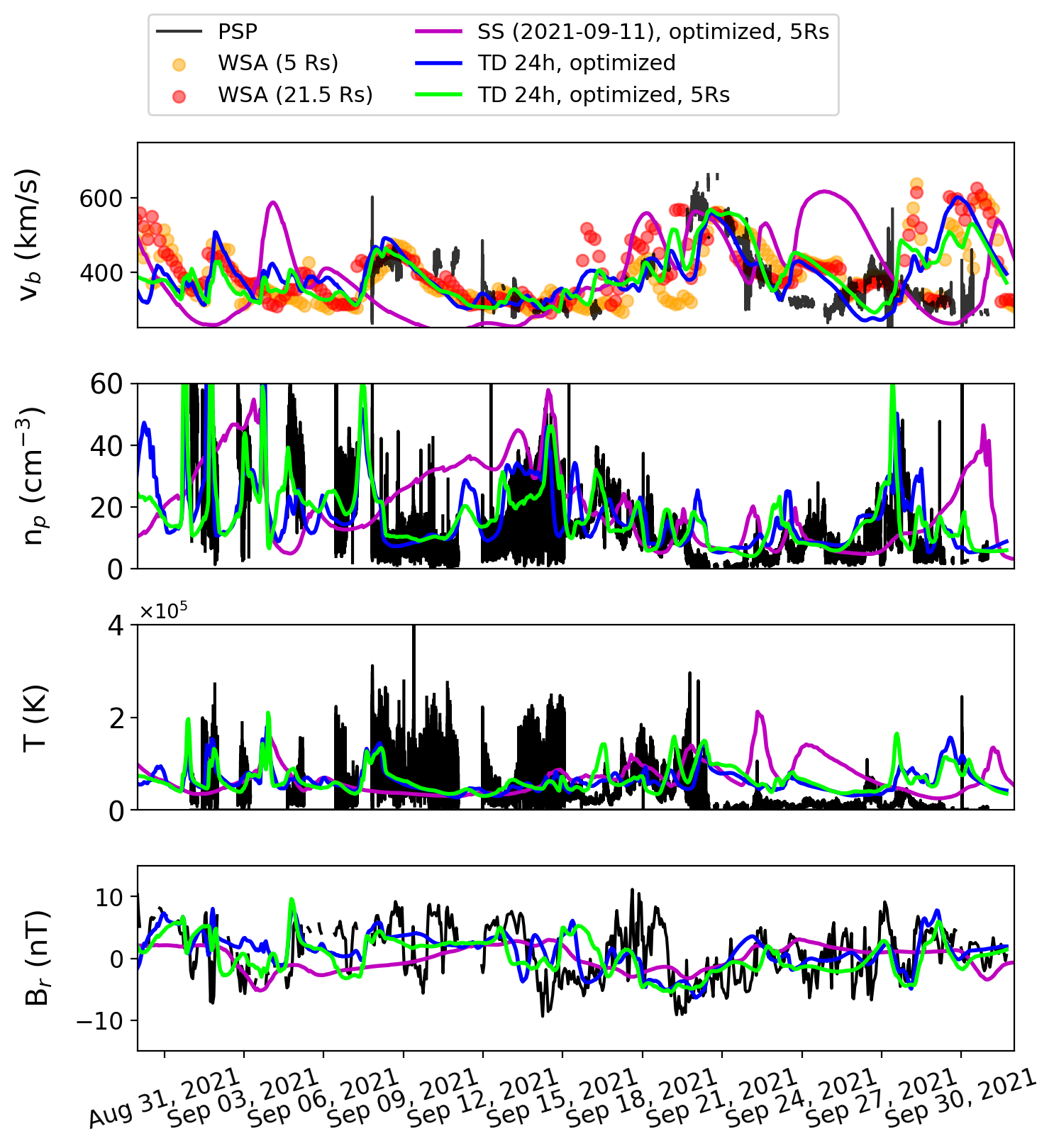}{0.7\textwidth}{}
          }        
\caption{Same as Figure~\ref{Fig:ACE_Sept2021} but for PSP.}
\label{Fig:PSP_Sept2021}
\end{figure}

\subsection{Other heliospheric locations} 

In Figures~\ref{Fig:ACE_Sept2021} and \ref{Fig:PSP_Sept2021}, we present comparisons of in situ observations and models' predictions at Earth and PSP, respectively, for the same time period as in Figures~\ref{Fig:STA_Sept2021} and \ref{Fig:STA_Sept2021_5Rs}. The OMNI data correspond to the combined merged hourly plasma and magnetic field (COHO1HR) observations obtained from CDAWeb (\url{https://cdaweb.gsfc.nasa.gov}) while the PSP data is a combination of the level 3 data from the Solar Probe Cup \citep[SPC;][]{SPC} and the Solar Probe Analyzers \citep[SPAN;][]{SPANi_1, SPANi_2} instruments. Similar to STEREO-A's location, the \mbox{WSA-GAMERA} modeling output with the optimized WSA velocity formula of eq.~\ref{WSA_vr_optimized} (blue time series) provide good agreement with the spacecraft measurements at Earth and PSP. The light green predictions that have been made not only with the optimized WSA velocity formula but also based on the WSA velocity maps created at 5~Rs (with radial magnetic fields scaled accordingly), are very similar to the blue time series and do not necessarily improve the predictions at Earth or PSP locations, opposite to the small but obvious improvements we showed for STEREO-A in Figure~\ref{Fig:STA_Sept2021_5Rs}. In some cases, these predictions also overestimate the density values at the stream interaction regions (see period around September~21-24,~2021) when the velocity transition from slow to fast solar wind is sharper. For comparison, magenta time series show the SS predictions according to both the optimized WSA formula of eq.~\ref{WSA_vr_optimized} and the WSA velocity maps created at 5~Rs. Discrepancies with observations are expected since we do not update the photospheric conditions on the Sun. Major discrepancies are noted for all parameters but especially for the radial magnetic field component between the SS predictions and observations. More specifically, the magenta time series show a quite flat and undisturbed Br pattern opposite to the TD simulations which highly vary in time, resembling the radial magnetic field observations much better.

\section{Evaluation of results: assessment of TD simulations during 2021}
\label{Evaluation2021}

In order to have a clearer idea which of the TD runs performs statistically better for a larger period of time, we ran one year of \mbox{WSA-GAMERA} simulations during 2021. Figures~\ref{Fig:Earth_FullYear2021_JantoJune} to \ref{Fig:PSP_FullYear2021_JultoDec} in Appendix~\ref{appendix:Appendix_A} show the default TD simulations (blue time series), the ones with the optimized WSA velocity formula (magenta time series) and those that have been produced with both the optimized WSA formula and WSA boundary conditions at 5~Rs instead of 21.5~Rs (light green time series). We have divided the yearly simulations into two 6-month intervals for each location of interest (Earth, STEREO-A, PSP) in order to facilitate the visual inspection of the time series for the reader, for all solar wind properties (bulk speed, density, temperature and radial magnetic field). Furthermore, we evaluate our results for bulk speed at Earth's and STEREO-A's locations according to three traditional error functions, namely, the mean absolute error (MAE), the root mean square error (RMSE) and the Pearson's correlation coefficient (PCC) as well as the dynamic time warping \citep[DTW;][]{Samara2022DTW}. Table~\ref{Table:ACE} and \ref{Table:STEREO-A} summarize the evaluation for the yearly predictions at the two heliospheric locations, respectively. Predictions at PSP were not evaluated because of large data gaps in measurements. 
CMEs were identified and removed from the observational datasets according to the Helio4Cast catalog found at \url{https://helioforecast.space/icmecat} \citep[see][for more details]{mostl2020}.

While it is straightforward to apply the traditional error functions for solar wind validation \citep[e.g., see][and references therein]{owens08, reiss16, bunting2024}, we will briefly describe the application of DTW which was more recently introduced for such purposes \citep[see][for more details]{Samara2022DTW}. First, we made sure that the output cadence of both observed and predicted time series was the same (1h) and that both time series consisted by the same number of elements (8761). Then we applied a 12-hour smoothing to the observed time series, namely, we took the arithmetic average of each element with its left and right 6 neighboring elements, respectively (otherwise known as running average). In this way, we reduced the small-scale fluctuations in the observed time series that unnecessarily penalize the cost of the evaluation process. Subsequently, we calculated the Sequence Similarity Factor (SSF) skill score, defined as follows: 

\begin{equation}
\centering
    \textrm{SSF} = \frac{\textrm{DTW}_{score}(O,M)}{\textrm{DTW}_{score}(O,B)}, \quad \textrm{SSF}\in [0,\infty).
\label{eq:SSF}
\end{equation}

\noindent where $DTW_{score}(O,M)$ is the cost of alignment between observations ($O$) and modeling output ($M$) while $DTW_{score}(O,B)$ is the cost of alignment between observations and the baseline model ($B$). As a baseline model, we used the climatological mean defined as the mean value of observations for 2021. The SSF is equal to zero when we have achieved the perfect forecast and equal to one when our forecast performs the same as the baseline model. For SSF values more than one, it means that the baseline models outperforms predictions. The closest the SSF is to 0, the best the forecast is.

\begin{table}[h!]
\centering
\renewcommand{\arraystretch}{1.5}
\begin{tabular}{|c |c |c |c|c|}
\hline 

\textbf{Earth} & TD 24h, default & TD 24h, optimized & TD 24h, optimized, 5~Rs & SS, optimized, 5~Rs\\ [0.5ex] 

\hline \hline
MAE & 85.0 & 72.0 & 68.0 & 96.0 \\
\hline 
RMSE & 112.0 & 97.0 & 93.0  & 125.0 \\
\hline 
PCC & 0.36 & 0.41 & 0.44  & 0.23 \\
\hline 
SSF & 0.85 & 0.61 & 0.58 & 0.82 \\
\hline \hline 
\end{tabular}
\caption{MAE, RMSE, PCC and SSF scores for the three TD runs with \mbox{WSA-GAMERA} for the whole year of 2021 at the location of Earth. The runs were performed by daily updating the magnetograms.}
\label{Table:ACE}
\end{table}

\begin{table}[h!]
\centering
\renewcommand{\arraystretch}{1.5}
\begin{tabular}{|c |c |c |c|c|}
\hline 

\textbf{STEREO-A} & TD 24h, default & TD 24h, optimized & TD 24h, optimized, 5~Rs & SS, optimized, 5~Rs\\ [0.5ex] 

\hline \hline
MAE & 108.0 & 88.0 & 78.0 & 95.0 \\
\hline 
RMSE & 139.0 & 117.0 & 107.0  & 126.0 \\
\hline 
PCC & 0.25 & 0.31 & 0.38  & 0.33 \\
\hline 
SSF & 1.1 & 0.73 & 0.59 & 0.79 \\
\hline \hline 
\end{tabular}
\caption{Same as Table~\ref{Table:ACE} but for STEREO-A.}
\label{Table:STEREO-A}
\end{table}

\begin{figure}[h!]
\centering
\gridline{\fig{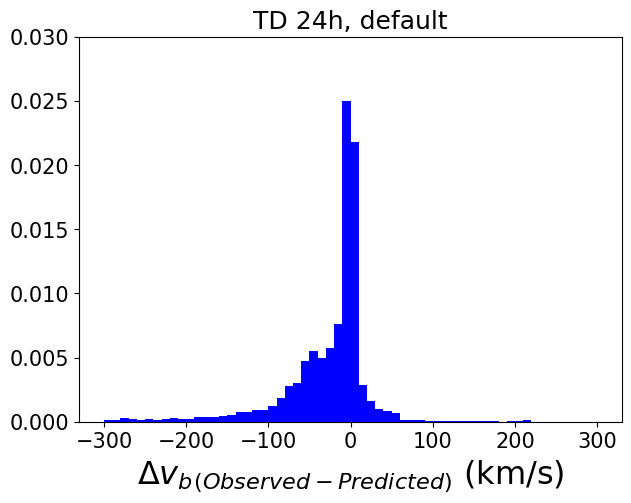}{0.3\textwidth}{}
            \fig{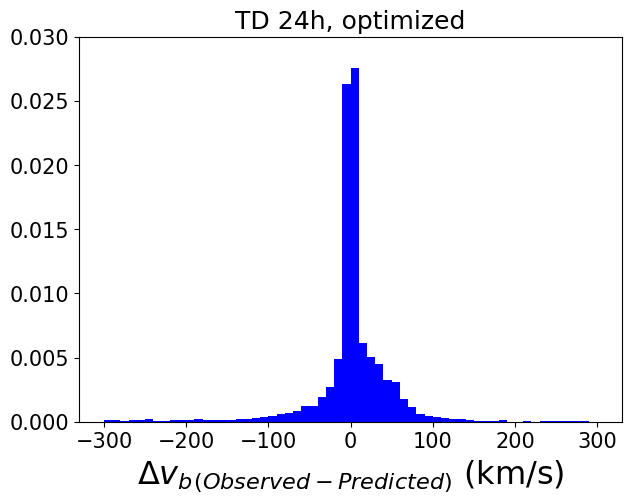}{0.3\textwidth}{}
            \fig{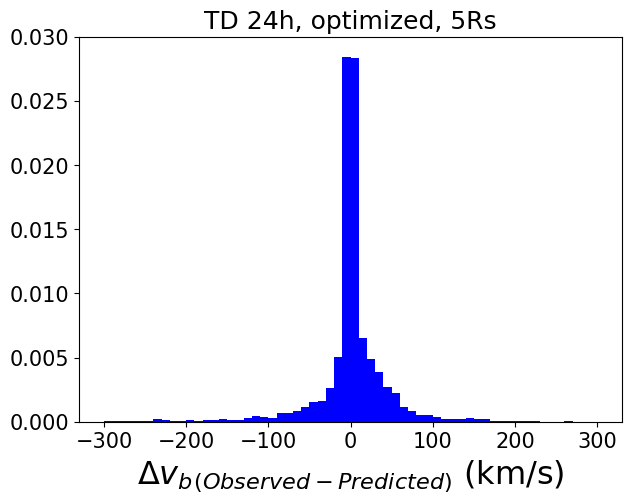}{0.3\textwidth}{}
          }     

\gridline{\fig{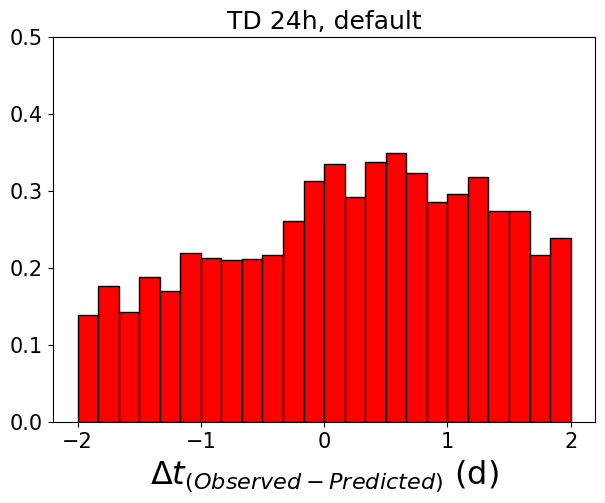}{0.3\textwidth}{}
            \fig{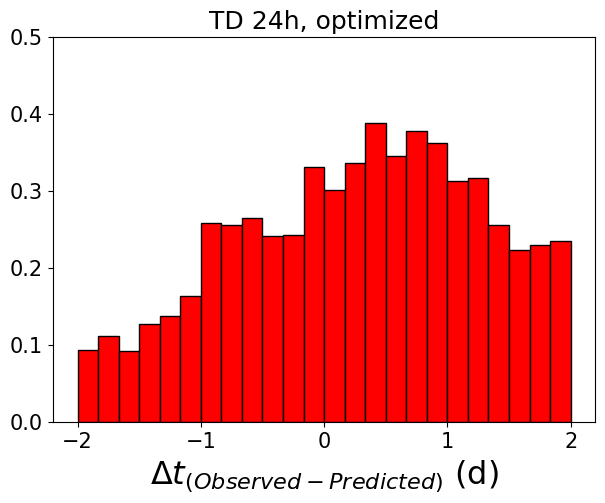}{0.3\textwidth}{}
            \fig{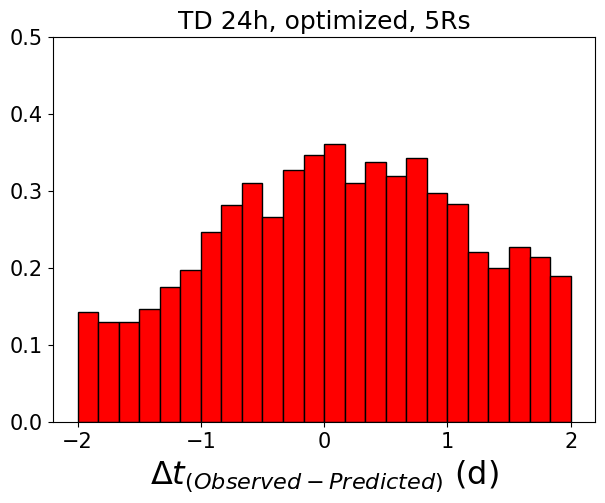}{0.3\textwidth}{}
          }       
\caption{Histograms of velocity differences (upper row) and time differences (lower row) between the aligned points in observed and modeled time series at Earth, as matched by DTW. The first column indicates the results of the default \mbox{WSA-GAMERA} TD run. The second column indicates the results from the \mbox{WSA-GAMERA} TD simulations that have been performed based on the optimized WSA velocity formula. The third column indicates the results from the \mbox{WSA-GAMERA} TD simulations that have been performed based on both the optimized WSA velocity formula and the WSA boundary conditions created at 5~Rs.}
\label{Fig:Histograms_DTW_Earth}
\end{figure}


\begin{figure}[h!]
\centering
\gridline{\fig{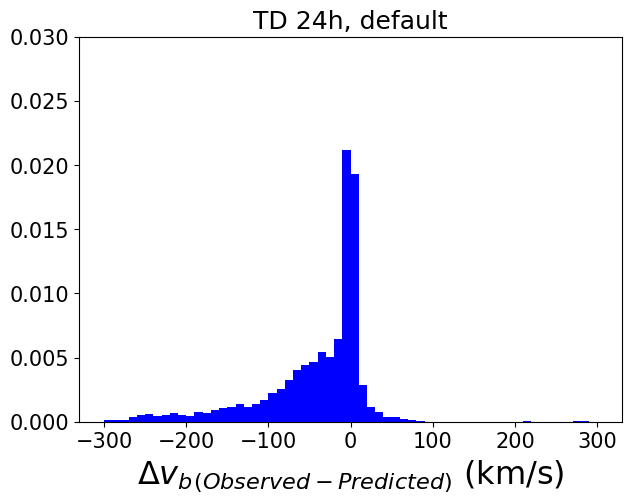}{0.3\textwidth}{}
            \fig{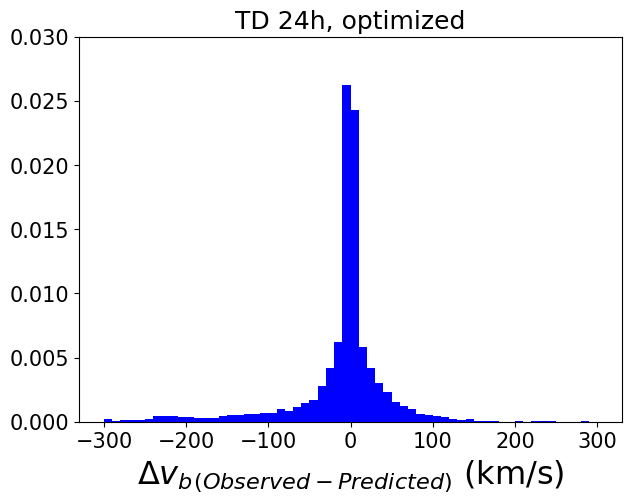}{0.3\textwidth}{}
            \fig{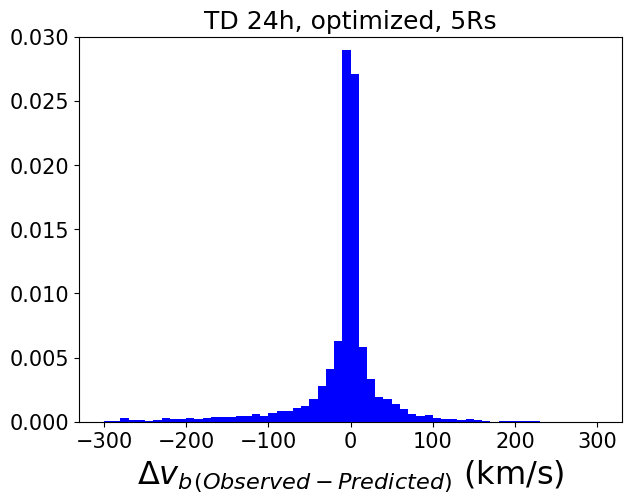}{0.3\textwidth}{}
          }    
          
\gridline{\fig{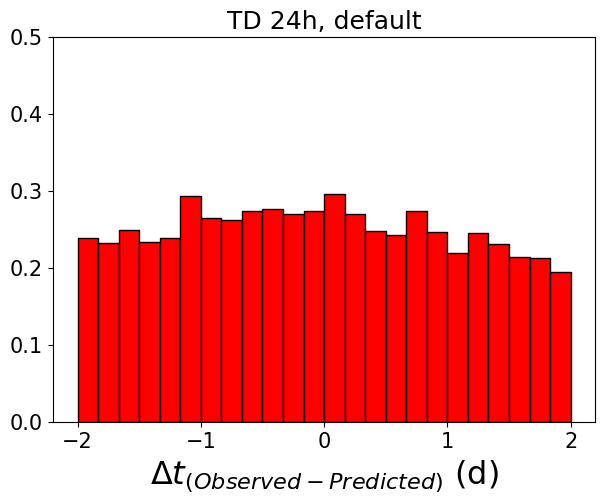}{0.3\textwidth}{}
            \fig{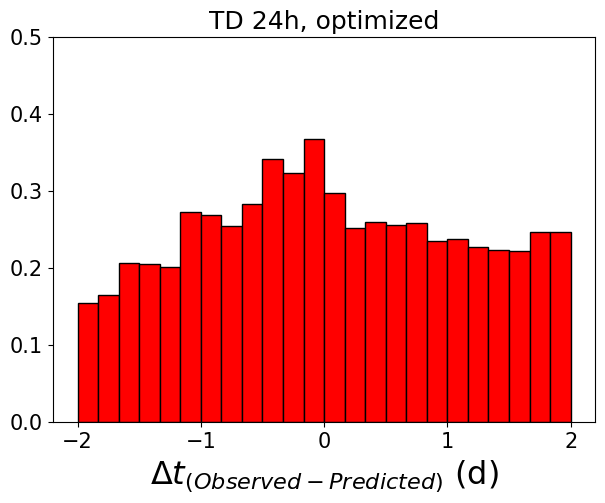}{0.3\textwidth}{}
            \fig{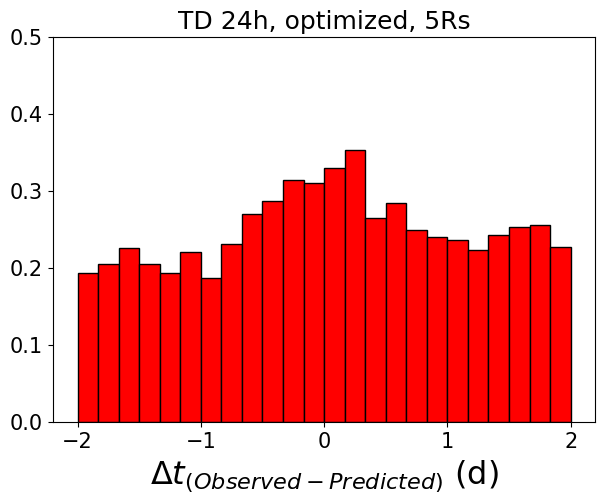}{0.3\textwidth}{}
          }     
\caption{Same as Figure~\ref{Fig:Histograms_DTW_Earth} but for STEREO-A.}
\label{Fig:Histograms_DTW_STA}
\end{figure}

Table~\ref{Table:ACE} shows that, for Earth location, the best TD simulations for 2021 are those that have been produced based on both the optimized WSA velocity formula and the WSA boundary conditions extracted at 5~Rs (instead of the usual height of 21.5~Rs), having scaled accordingly the radial magnetic field. This is confirmed from all metrics we considered (lowest MAE, RMSE, SSF and highest PCC among all TD runs). The second best performance comes from the simulations that have been conducted with the optimized WSA velocity formula (eq.~\ref{WSA_vr_optimized}). Both of those runs outperform the default \mbox{WSA-GAMERA} output which is characterized by the highest MAE, RMSE, SSF as well as the lowest PCC. The same pattern is followed for the predictions at STEREO-A location (Table~\ref{Table:STEREO-A}). Namely, the best TD simulations for 2021 are those that have been produced based on both the optimized WSA velocity formula and the WSA boundary conditions extracted at 5~Rs instead of the usual height of 21.5~Rs (lowest MAE, RMSE, SSF and highest PCC among all TD runs). The second best performance comes from the simulations that have been conducted with the optimized WSA velocity formula, and the third best performance from the default \mbox{WSA-GAMERA} output. In the latter case, the SSF value is more than one showing that the baseline model (climatological mean) performed better than the TD simulations. 

Figures~\ref{Fig:Histograms_DTW_Earth} and \ref{Fig:Histograms_DTW_STA} show the histograms of velocity and time differences between the aligned points of the observed and modeled time series as matched by DTW at Earth and STEREO-A, respectively, for 2021. It would be too complex to show the exact DTW alignments for the whole year of 2021, therefore, we present an example of those in Figure~\ref{Fig:DTW_alignments} of the Appendix~\ref{appendix:Appendix_A} only for the period between September~7~-~October~6,~2021. Green lines indicate which points of one time series better correspond to points from the other time series. Based on those alignments, the histograms of velocity and time differences can be created, similarly to Figures~\ref{Fig:Histograms_DTW_Earth} and \ref{Fig:Histograms_DTW_STA}. Based on the latter plots, for the full year of 2021 we notice a systematic overestimation of the predicted speeds from the default TD run at both heliospheric locations (see $\Delta$v $<$ 0 regime in the upper left panels of Figures~\ref{Fig:Histograms_DTW_Earth} and \ref{Fig:Histograms_DTW_STA}). This overestimation of modeled speeds compared to measurements is much eliminated in the TD runs with the optimized WSA velocity formula (upper middle plots of Figures~\ref{Fig:Histograms_DTW_Earth} and \ref{Fig:Histograms_DTW_STA}), and even more eliminated in the TD runs with both the optimized WSA velocity formula and the WSA boundary created at 5~Rs (upper right plots of Figures~\ref{Fig:Histograms_DTW_Earth} and \ref{Fig:Histograms_DTW_STA}). Regarding the histograms of time differences, we notice that the maximum difference between observations and predictions is between $\pm$2 days in all cases. This is because we imposed a priori a time window of $\pm$2 days within which DTW was allowed to do the alignments between the points of the two sequences. The time-differences become increasingly smaller as we go from the default \mbox{WSA-GAMERA} simulations (lower left plots of Figures~\ref{Fig:Histograms_DTW_Earth} and \ref{Fig:Histograms_DTW_STA}), to the optimized \mbox{WSA-GAMERA} output (lower middle plots of Figures~\ref{Fig:Histograms_DTW_Earth} and \ref{Fig:Histograms_DTW_STA}), to the \mbox{WSA-GAMERA} output with both the optimized WSA velocity formula and the WSA boundary conditions created at 5~Rs (lower right plots of Figures~\ref{Fig:Histograms_DTW_Earth} and \ref{Fig:Histograms_DTW_STA}). This is more noticeable for the Earth (Figures~\ref{Fig:Histograms_DTW_Earth}) rather than the STEREO-A \ref{Fig:Histograms_DTW_STA} case.

Since all metrics showed that the best TD simulations for 2021 are those that have been produced based on both the optimized WSA velocity formula and the WSA boundary conditions generated at 5~Rs instead of the usual height of 21.5~Rs (this is also visible by eye in 
Figures~\ref{Fig:Earth_FullYear2021_JantoJune} to \ref{Fig:STA_FullYear2021_JultoDec} in Appendix~\ref{appendix:Appendix_A}), we further perform yearly SS \mbox{WSA-GAMERA} simulations based on this approach. The output of the SS simulations for 2021 is shown in Figures~\ref{Fig:Earth_FullYear2021_JantoJune} to \ref{Fig:PSP_FullYear2021_JultoDec} in Appendix~\ref{appendix:Appendix_A} (red time series) so that the comparison with the rest TD simulations is visible. To perform the SS runs, we picked an input map every 26 calendar days. The selected date of this map was at the middle of this 26-days period so in total we performed 13 SS \mbox{WSA-GAMERA} simulations and combined them by using a rolling moving average smoothing. The SS results were also assessed by the different metrics at Earth's and STEREO-A's locations (see last column of Table~\ref{Table:ACE} and \ref{Table:STEREO-A}). More specifically, at Earth location, the SS simulations perform worse than any TD run according to the traditional error functions (MAE, RMSE, PCC) and slightly better than the TD 24h default run according to DTW. For STEREO-A location, the MAE, RMSE, and SSF metrics show that the SS runs perform a bit better than the 24h default run and worse than the TD simulations with the optimized WSA velocity formula. Only the PCC shows a slightly different behavior for SS simulations, namely, the fact that they slightly outperform the TD runs with the optimized WSA velocity formula. Overall, from Table~\ref{Table:ACE} and \ref{Table:STEREO-A} we see that the fully optimized SS simulations at no case outperform the fully optimized TD simulations (comparison between 3rd and 4th columns).

\section{Summary and conclusions}
\label{section:Summary&Conclusions}

In this study, we presented the first GAMERA TD MHD simulations of the solar wind in the inner heliosphere. To approximate the time-evolution of the solar corona, we daily updated the inner boundary of the GAMERA heliospheric domain at 21.5$~Rs$ according to WSA maps which are produced from a series of GONG ADAPT magnetograms (first realization was selected). We initially compared SS and TD solar wind simulations from the \mbox{WSA-GAMERA} model with in situ data and WSA 1D kinamatic predictions at STEREO-A's location for the whole month of September~2021. We noticed that, even though TD simulations provided better results compared to SS ones, their performance was still far from optimal when compared to observations and the WSA 1D kinematic model. In many instances, TD predictions were inconsistently predicting HSSs, or overestimating the slow solar wind, or wrongly capturing the transitions between the slow and fast solar wind, so we had to carefully explore the reasons why this was the case.

The factors that caused discrepancies between the modeling results and observations mainly had to do with the input solar wind boundary conditions to the GAMERA model. More specifically, big discrepancies occured because the WSA velocity formula is not properly calibrated for use by heliospheric MHD models, and also because the distance from the Sun at which we extract the WSA boundary conditions is too high. Besides these two factors, we are unable to know the phenomena that take place in the invisible side of the Sun (such as active regions) whose sudden and partial appearance from the solar east limb introduces significant uncertainty in predicting the solar wind farther out. We found that all of these reasons have a significant contribution to the accuracy and consistency of modeling the solar wind parameters in various locations in the heliosphere (Earth, STEREO-A, PSP). 

Since it is currently impossible to account for the latter factor, we focused on the other two. In Section~\ref{Section:Why do models get it wrong} we presented how the default WSA velocity formula (eq.~\ref{WSA_vr}) can be optimized for implementation into MHD models (eq.~\ref{WSA_vr_optimized}) in order to give better predictions at any point of interest in the heliosphere. We also showed how the MHD solar wind simulations are affected by the height at which we extract the WSA boundary conditions (comparison between the traditional outer coronal radius of 21.5~Rs that is usually used with MHD models vs 5~Rs which is used with the 1D kinematic WSA model). Our results showed that the MHD solar wind simulations performed with boundary conditions extracted at 5~Rs (instead of 21.5~Rs), provided better agreement with the observations in situ. In doing that, it is important to emphasize that we did not change the inner boundary of the GAMERA model from 21.5~Rs to 5~Rs; we just generated the WSA conditions at 5~Rs and we gave them as input to 21.5~Rs assuming negligible acceleration of the solar wind and scaling accordingly the magnetic field. During this process, the 1D kinematic WSA model which ballistically propagates the solar wind from the outer coronal boundary (5~Rs) outward, helped us confirm our results. Namely, working vice versa, we verified that if we change the height at which the WSA boundary conditions are extracted for use with the ballistic 1D kinematic propagation (from 5~Rs to 21.5~Rs, in this case), we get very similar results to the MHD simulations when we employ the outer coronal radius of 21.5~Rs. This is a clear evidence of the significance and influence of the boundary conditions driving a heliospheric model that are much more important than the details/physics of the heliospheric model itself. It is also useful to highlight that no longitudinal shift was applied when working with coronal output from 5~Rs to 21.5~Rs. This is because, based on \citet[][]{macneil2022}, the assumption of coronal co-rotation out to 21.5~Rs~au is roughly canceled out by the assumption of little solar wind acceleration between 21.5~Rs and 1~au.

To validate our results, in Section~\ref{Evaluation2021} we performed TD runs for the whole year of 2021. We compared the bulk speed predictions from three types of \mbox{WSA-GAMERA} runs: (a) default TD simulations, (b) TD simulations with the optimized WSA velocity formula of (eq.~\ref{WSA_vr_optimized}) and (c) TD simulations with both the optimized WSA velocity formula of (eq.~\ref{WSA_vr_optimized}) and WSA boundary conditions that were created at 5~Rs, instead of the usual 21.5~Rs. All such TD runs were performed by daily updating GONG ADAPT magnetograms. We assessed the results with a number of traditional error functions (i.e., MAE, RMSE, PCC) as well as the DTW method, and found that the best performance for 2021 was given by the latter (c) run. The second best performance was from run (b). Both runs (b) and (c) outperformed run (a), namely, the default \mbox{WSA-GAMERA} TD simulations. Our results are valid for both Earth and STEREO-A locations.

Last but not least, we performed SS simulations for the whole year of 2021 following the same approach with run (c). Evaluation of the SS output with observations at Earth's location indicated that SS runs performed worse than all TD runs according to the traditional error functions, and the same as or slightly better than the default TD output based on DTW. For STEREO-A location, they performed better than the TD default runs and worse than the TD runs with the optimized velocity formula. However, they could not outperform the TD simulations that were based on both the optimized WSA velocity formula and the WSA boundary conditions extracted at 5~Rs. It is important to note that, with a closer look at Figures~\ref{Fig:Earth_FullYear2021_JantoJune} to \ref{Fig:STA_FullYear2021_JultoDec} in Appendix~\ref{appendix:Appendix_A}, one can identify periods during which the SS simulations performed better than any of the TD output. This can occasionally happen when the photospheric input map is good enough and has the necessary information that can lead to good solar wind predictions overall. However, this is a quite random phenomenon and we should not deduce that the SS approach should be preferred compared to the TD one as the input maps in that case certainly do not contain the most updated photospheric magnetic field information.

Our findings indicate that even though TD solar wind simulations provide improved predictions compared to the traditional SS ones (shown also by \citet[][]{merkin2016time, linker2016empirically, owens2024}) as they are being driven by boundary conditions that are more timely and accurately representing the solar corona, we should not expect faultless output from them. Modelers whom work with similar modeling pipelines as ours (namely, with a combination of a semi-empirical coronal model, like WSA, and a 3D MHD heliospheric model, like GAMERA) should consider using a calibrated version of the WSA velocity formula when WSA is coupled to MHD models. This is because, the WSA formula used by the traditional WSA model is only fine-tuned appropriately to provide well modeled speeds at 1~au according to a 1D ballistic (and not MHD) approach. The calibration we performed in the frame of this study was based on the methodology of \citet[][]{Samara2024} which relied on PSP data from the first eight encounters. More specifically, we changed the value of some constant parameters in eq.~\ref{WSA_vr} that could be fine-tuned according to observations and past studies. However, a complete reformulation of this relationship with more advanced techniques (e.g., artificial intelligence) and more data from current and upcoming missions (PSP, Solar Orbiter, PUNCH) could potentially lead to even better results. Besides the calibration of the WSA velocity formula, modelers should keep in mind that the SCS model can frequently create an overly flat structure for the HCS that can lead to modeling discrepancies in heliospheric simulations. Changing the height at which WSA boundary maps are being created, changes the outer radius of the SCS model and adjusts the thickness of the HCS. Whether this thickness is systematically different between solar minimum and maximum, is left to be explored.

\begin{acknowledgments}
Resources supporting this work were provided by the NASA High-End Computing (HEC) Program through the NASA Advanced Supercomputing (NAS) Division at Ames Research Center. E.S. research was supported by an appointment to the NASA Postdoctoral Program at the NASA Goddard Space Flight Center, administered by Oak Ridge Associated Universities under contract with NASA. C.N.A is supported by the NASA competed Heliophysics Internal Scientist Funding Model (ISFM). E.P. and A.M. were supported by the AFOSR YIP program award number FA9550-21-1-0276. V.G.M. was supported by the AFOSR grant FA9550-21-1-0457.
\end{acknowledgments}

\bibliography{bibliography}{}
\bibliographystyle{aasjournal}


\clearpage

\begin{appendices}
\section{Yearly TD predictions at Earth for 2021}
\label{appendix:Appendix_A}

\begin{figure}[h!]
\centering
\gridline{\fig{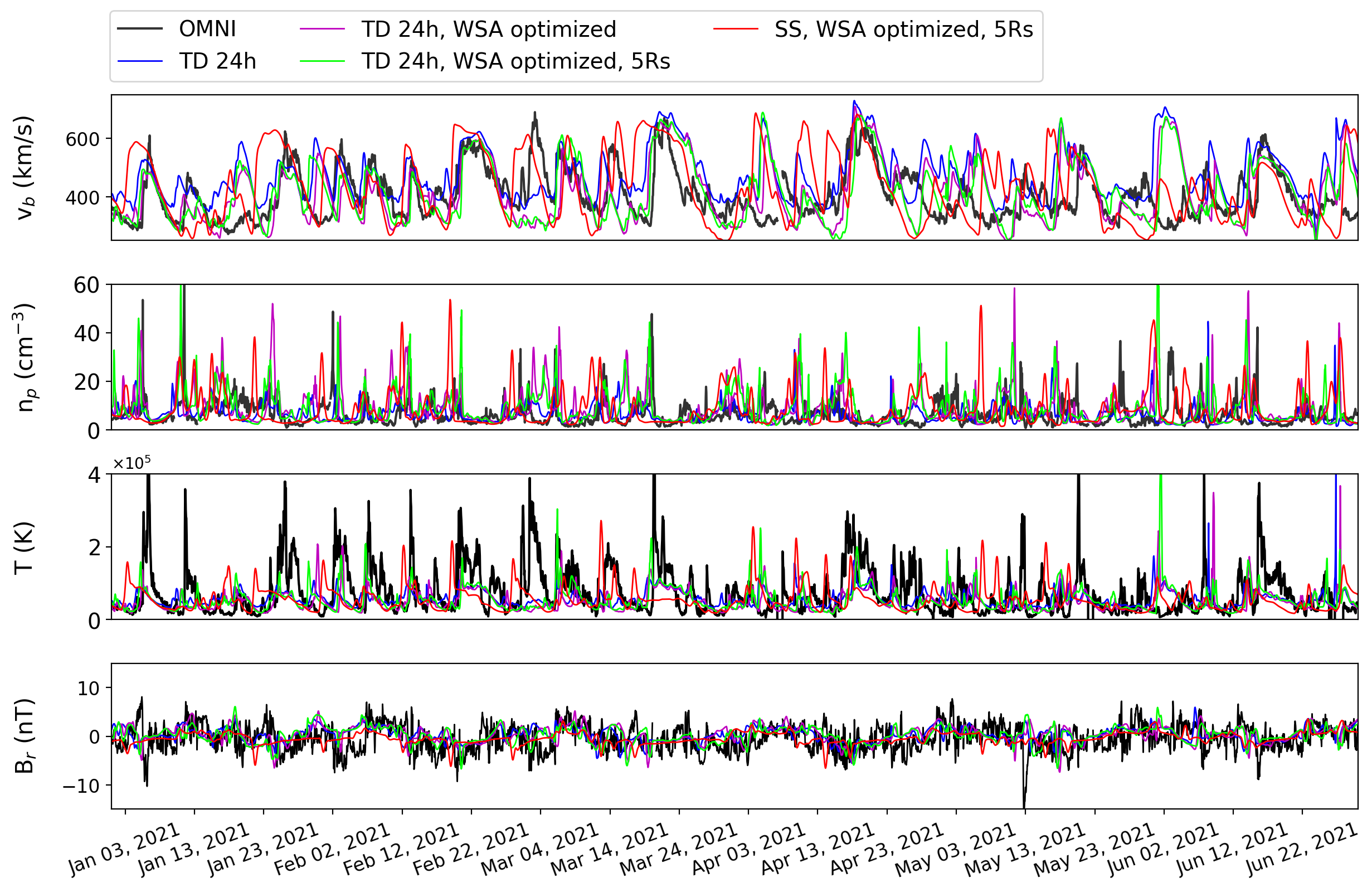}{0.99\textwidth}{} }        
\caption{Comparison of in situ solar wind observations (black) with TD and SS predictions from \mbox{WSA-GAMERA} between January-June, 2021. Blue color indicates the results of the default TD run, magenta color shows the results of the TD run with the optimized WSA velocity formula, while for the light-green time series we have employed the WSA boundary conditions at 5~Rs on top of the optimized WSA velocity formula. The updating frequency cadence of the magnetograms is daily. The SS simulations are shown in red.}
\label{Fig:Earth_FullYear2021_JantoJune}
\end{figure}

\begin{figure}[h!]
\centering
\gridline{\fig{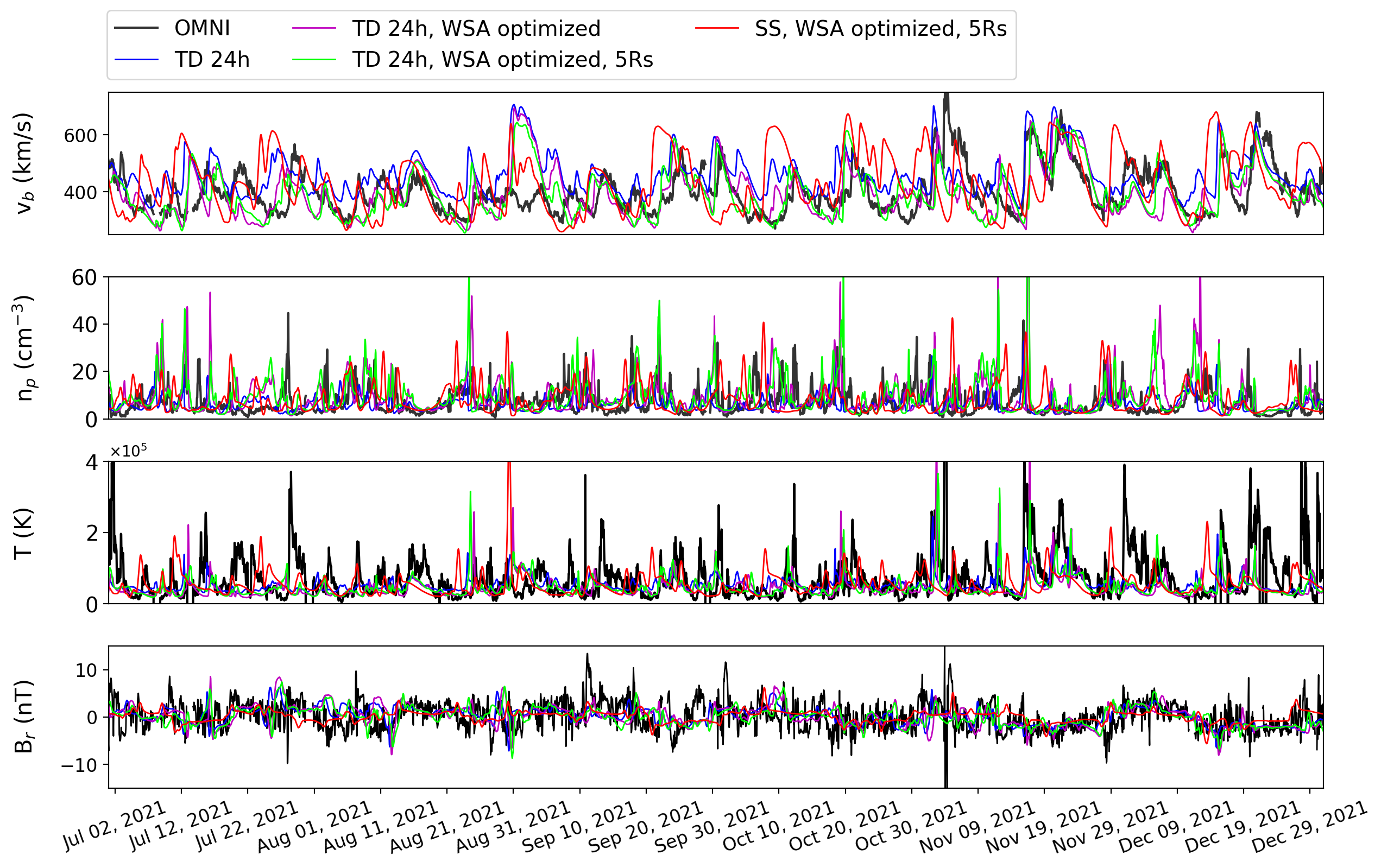}{0.99\textwidth}{}}        
\caption{Same as Figure~\ref{Fig:Earth_FullYear2021_JantoJune} but for the time interval July-December, 2021. }
\label{Fig:Earth_FullYear2021_JultoDec}
\end{figure}

\begin{figure}[h!]
\centering
\gridline{\fig{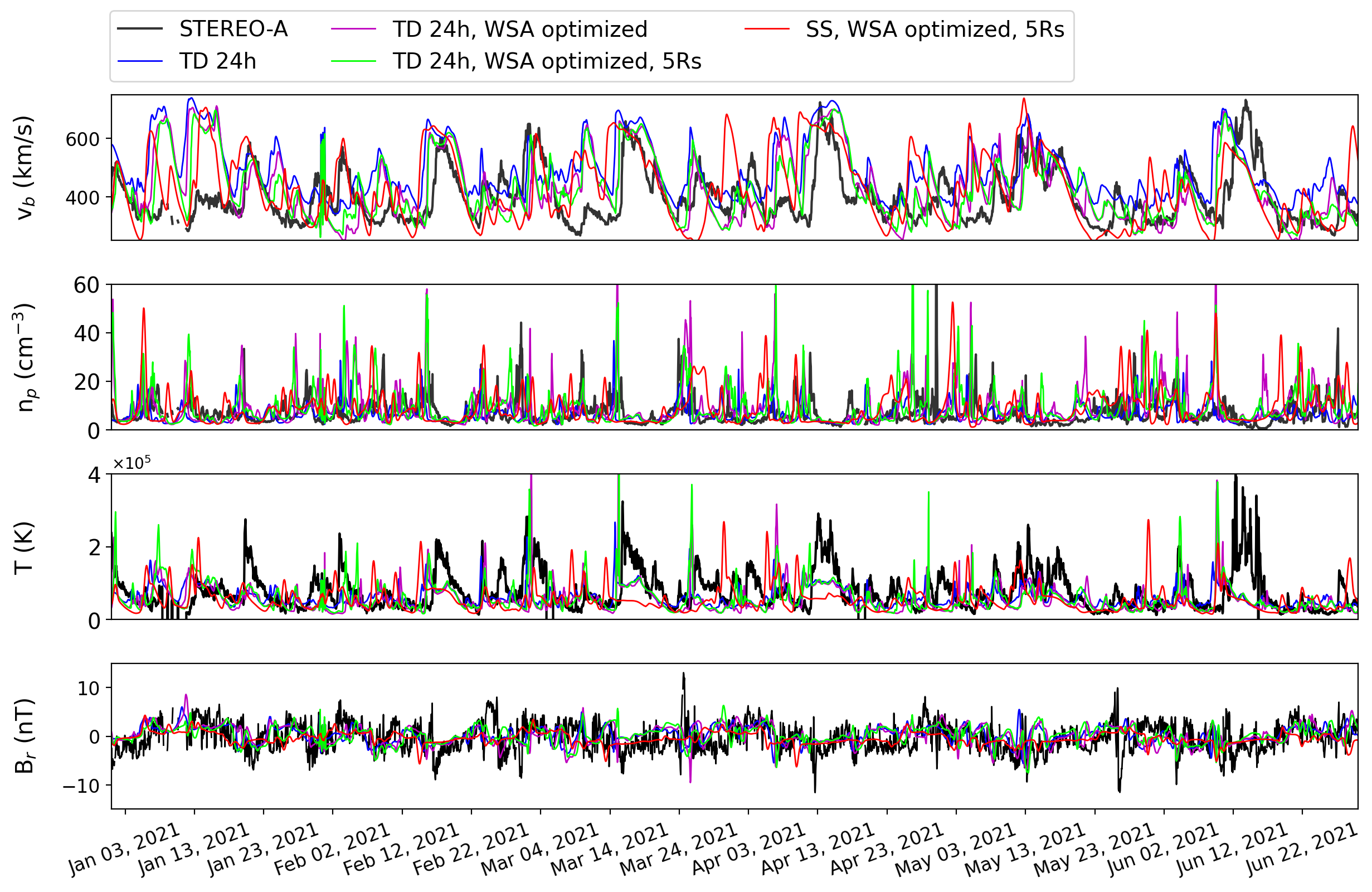}{0.99\textwidth}{}}        
\caption{Same as Figure~\ref{Fig:Earth_FullYear2021_JantoJune} but for STEREO-A location. }
\label{Fig:STA_FullYear2021_JantoJune}
\end{figure}

\begin{figure}[h!]
\centering
\gridline{\fig{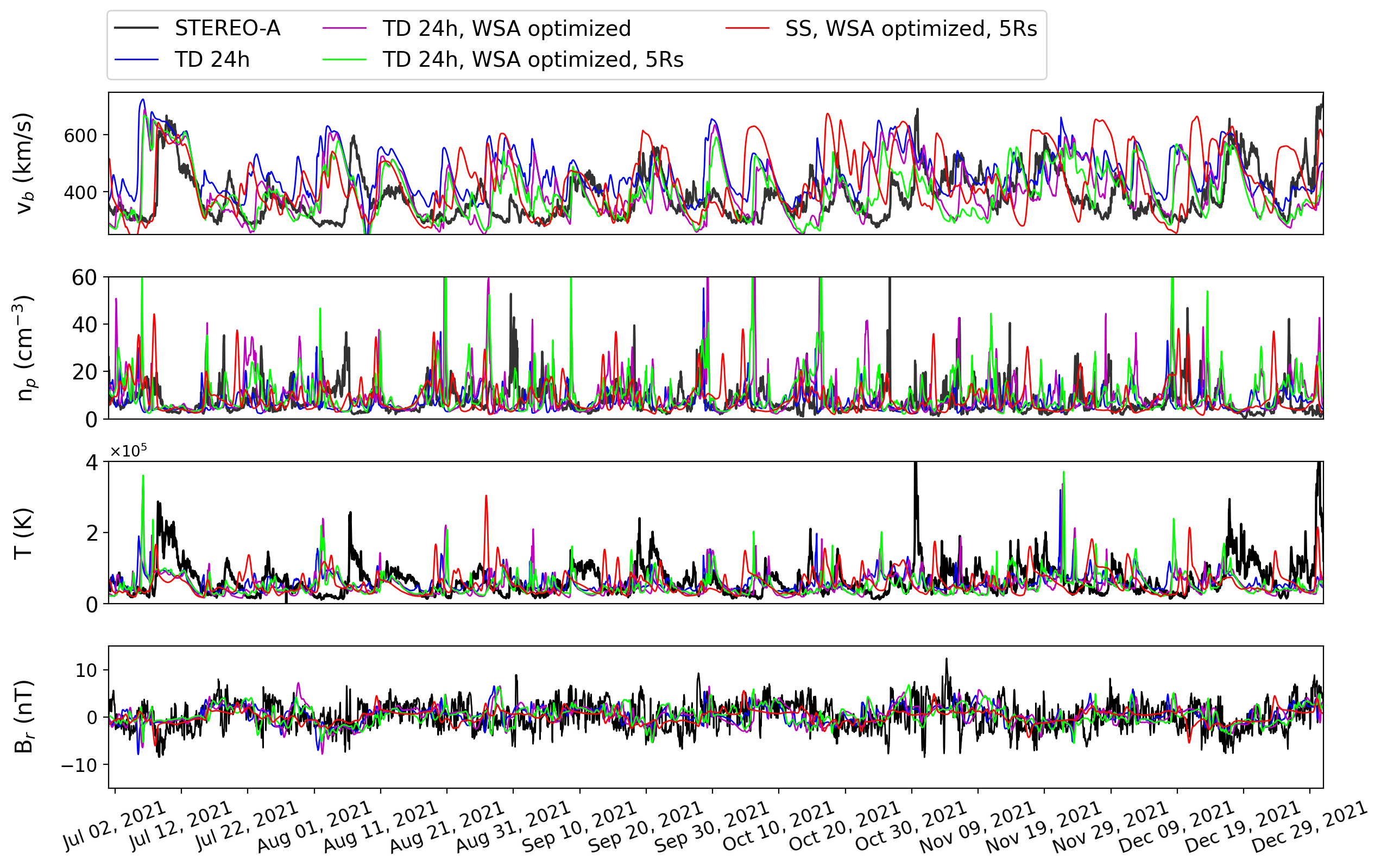}{0.99\textwidth}{}}        
\caption{Same as Figure~\ref{Fig:Earth_FullYear2021_JultoDec} but for STEREO-A location. }
\label{Fig:STA_FullYear2021_JultoDec}
\end{figure}

\begin{figure}[h!]
\centering
\gridline{\fig{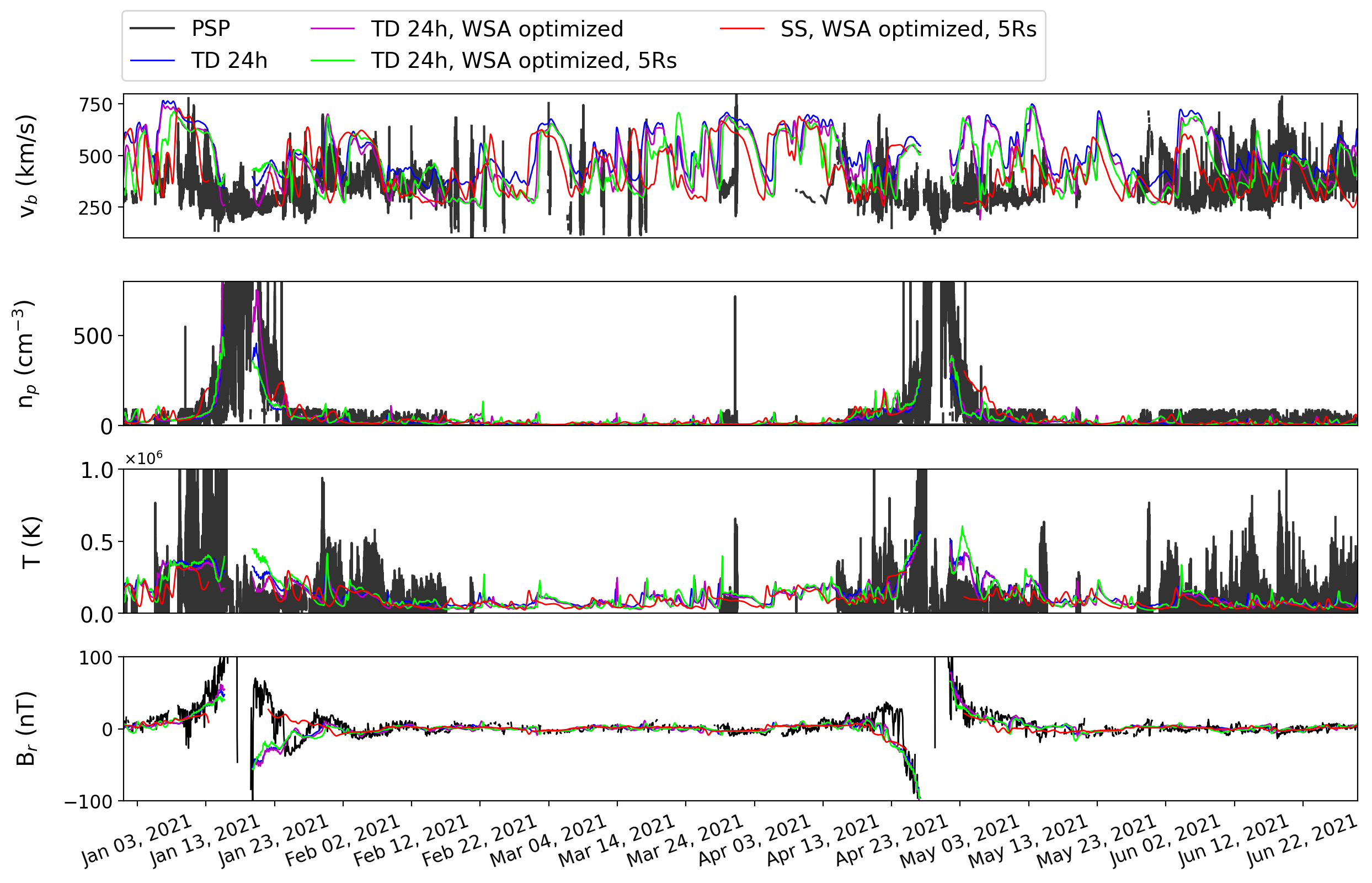}{0.99\textwidth}{}}        
\caption{Same as Figure~\ref{Fig:Earth_FullYear2021_JantoJune} but for PSP location. Gaps on the modeling output indicate intervals at which PSP is very close to the inner heliospheric boundary (21.5~Rs). }
\label{Fig:PSP_FullYear2021_JantoJune}
\end{figure}

\begin{figure}[h!]
\centering
\gridline{\fig{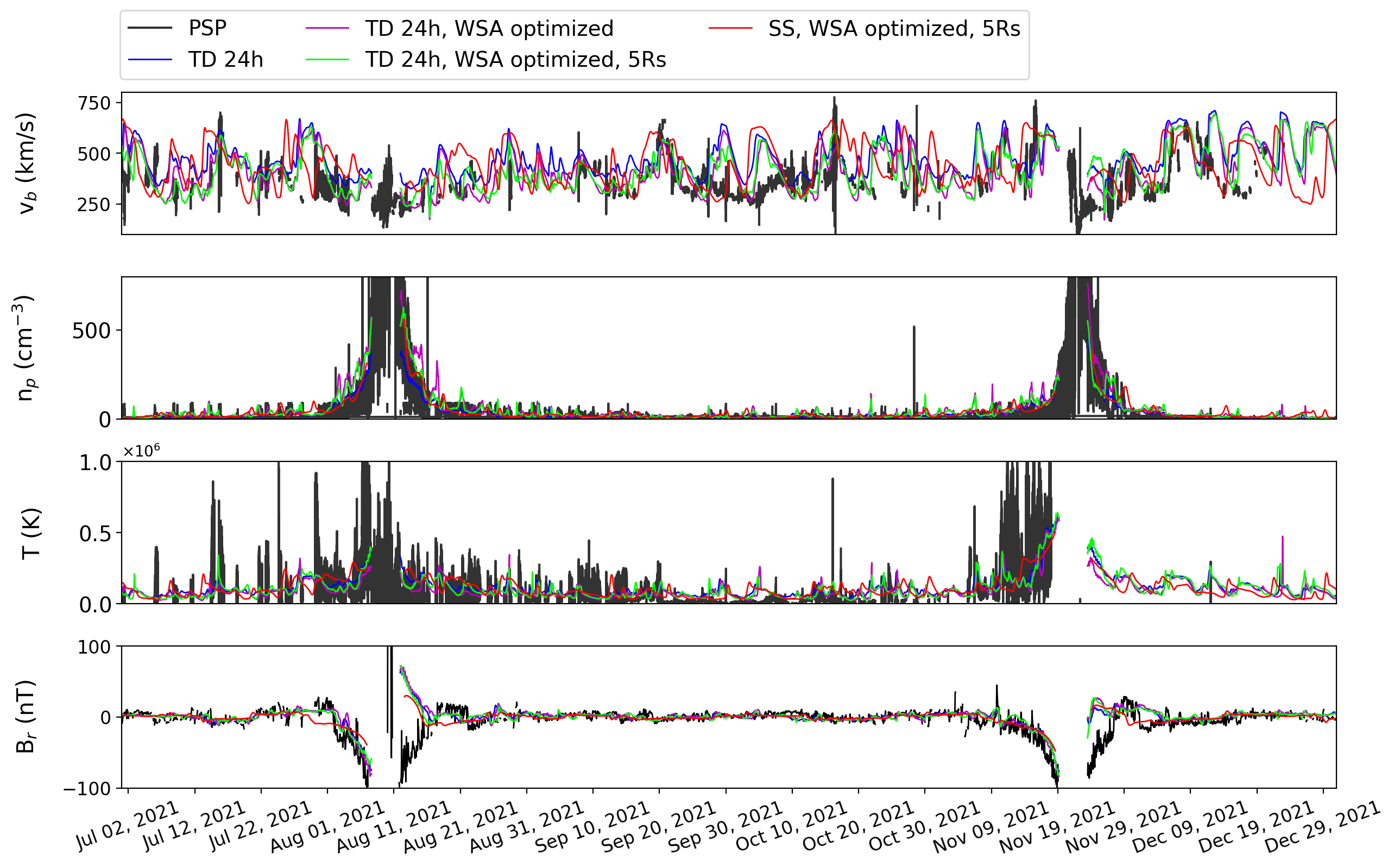}{0.99\textwidth}{}}        
\caption{Same as Figure~\ref{Fig:Earth_FullYear2021_JultoDec} but for PSP location. Gaps on the modeling output indicate intervals at which PSP is very close to the inner heliospheric boundary (21.5~Rs).}
\label{Fig:PSP_FullYear2021_JultoDec}
\end{figure}

Figures~\ref{Fig:Earth_FullYear2021_JantoJune} to \ref{Fig:PSP_FullYear2021_JultoDec} show the comparison between observations (in black), SS \mbox{WSA-GAMERA} simulations (in red) and TD \mbox{WSA-GAMERA} simulations based on three different runs. The first run (in blue) indicates the default TD simulation. The second run (in magenta) shows the TD simulation with the optimized WSA velocity formula of eq.~\ref{WSA_vr_optimized}. The third run (in light-green) shows the TD simulation which, in addition to the optimized WSA velocity formula, has been generated by employing the WSA velocity boundary map created at 5~Rs instead of the usual 21.5~Rs. In the latter case, we highlight the fact that the boundary between the coronal and heliospheric modeling domain did not change but stayed as is, at 21.5~Rs. However, instead of generating the WSA velocity maps at that distance, we generated it instead at 5~Rs and we provided them as input to 21.5~Rs, having scaled the radial magnetic field accordingly. All TD simulations presented at these plots have been produced according to daily magnetograms update.

\begin{figure}[h!]
\centering
\gridline{\fig{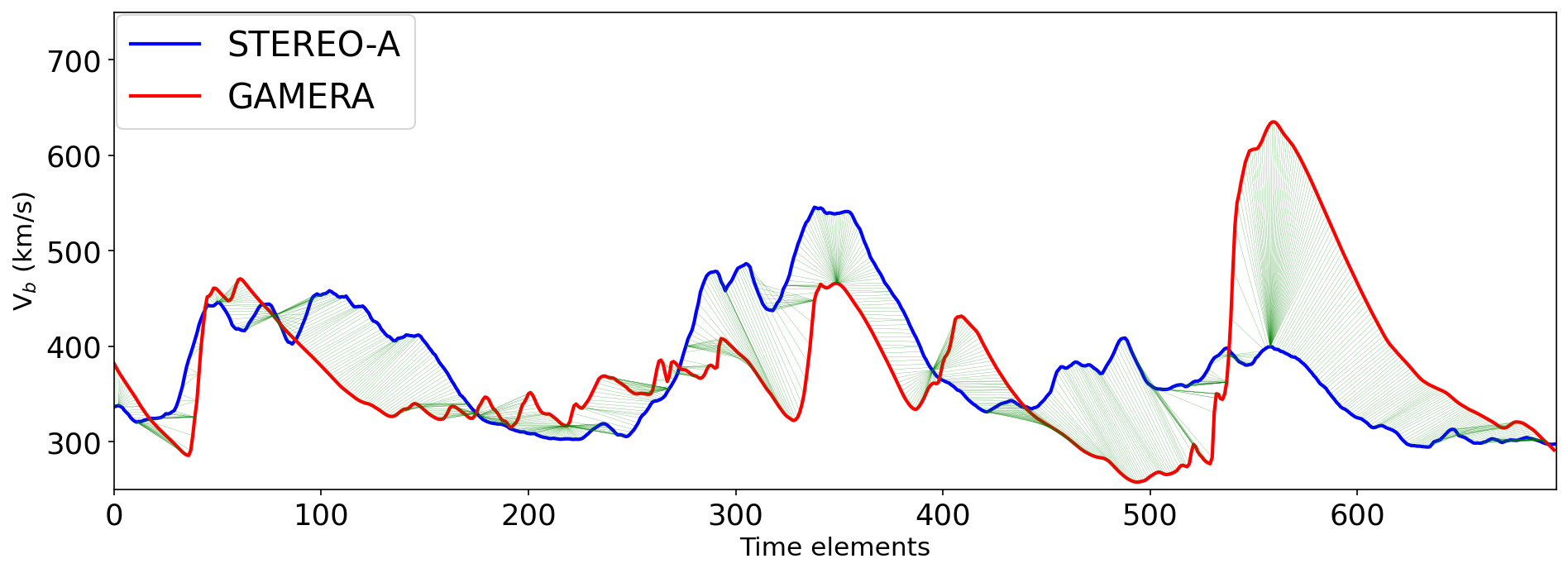}{0.99\textwidth}{} }  
\caption{Example of how DTW is applied between observed and modeled time series at STEREO-A location between September 7-October 6, 2021. The green lines show the alignments between the points from one timeseries that better correspond to the points of the other time series according to DTW. Based on these alignments, the histograms of velocity and time differences can be created.}
\label{Fig:DTW_alignments}
\end{figure}

\end{appendices}

\end{document}